\documentclass[%
 reprint,
 jmp,
superscriptaddress,
 amsmath,amssymb,
 aps,
]{revtex4-2}

\usepackage{graphicx}% Include figure files

\usepackage{dcolumn}% Align table columns on decimal point
\usepackage{bm}% bold math
\usepackage{mathtools}
\usepackage{braket}
\usepackage{xcolor}
\begin{document}

\preprint{APS/123-QED}

\title{Time-resolved characterization of pulsed squeezed light from a strongly driven silicon nitride microresonator}

\author{Emanuele Brusaschi}
\email{corresponding author:emanuele.brusaschi01@universitadipavia.it} 
\author{Marco Liscidini}
\affiliation{Dipartimento di Fisica, Università di Pavia, Via Bassi 6, 27100 Pavia, Italy.}
\author{Matteo Galli}
\affiliation{Dipartimento di Fisica, Università di Pavia, Via Bassi 6, 27100 Pavia, Italy.}
\author{Daniele Bajoni}
\affiliation{Dipartimento di Ingegneria Industriale e dell'Informazione, Università di Pavia, Via Ferrata 5, 27100 Pavia, Italy.}
\author{Massimo Borghi}
\affiliation{Dipartimento di Fisica, Università di Pavia, Via Bassi 6, 27100 Pavia, Italy.}

\date{\today}% It is always \today, today,

\begin{abstract}
Silicon nitride microresonators driven by strong pump pulses can generate squeezed light in a dominant spectral-temporal mode, a central resource for continuous-variable quantum computation. In the high parametric gain regime, several effects, including self- and cross-phase modulation as well as time-ordering corrections, become significant and can degrade source performance.\\ In this work, we comprehensively investigate the generation of squeezed light from a silicon nitride resonator under pulsed pumping, spanning from low to high parametric gain up to $\sim16$ photons/pulse. We experimentally study how the average photon number and the first- and second-order correlations of the squeezed marginal modes evolve with increasing pulse energy, across various frequency detunings and pulse durations.\\
Furthermore, we analyze the errors introduced by multi-pair emissions in estimating the joint temporal intensity via time-resolved coincidence measurements. We propose and demonstrate an error-correction strategy based on the marginal distributions of time-resolved multi-photon events.\\
%We found that optimal conditions for achieving both high gain and single-mode emission are achievied with detunings in the order of the linewidth of the resonance.
Our results provide a practical strategy for optimizing the gain and the temporal mode structure of pulsed squeezed light sources in microresonators, elucidating the physical mechanisms and limitations that govern source performance in the high gain regime. 
\end{abstract}

\maketitle

%\tableofcontents

\section{\label{sec:intro} Introduction}
The generation of squeezed light is one of the cornerstones of fundamental quantum science, and a key enabler for emerging quantum information technologies \cite{squeezed_light_30_years_review, GaussianQuantumInfo}.
Materials with a strong third-order nonlinearity are widely used to produce either single or two-mode squeezed light by spontaneous four-wave mixing (SFWM) through the annihilation of pairs of pump photons from a bright coherent beam and the creation of time-energy correlated photon pairs. %, generating entangled photon pairs commonly referred to as the ‘signal’ and ‘idler’, can be generated through spontaneous four-wave mixing (SFWM).
As quantum information processing grows in size and complexity, the need for scalable and integrated squeezed light sources becomes increasingly critical \cite{aghaee2025scaling,psiquantum2025manufacturable}. Among them, microring resonators offer an exclusive combination of compact footprint, strong field enhancement and narrow-band photon emission \cite{caspani2017integrated}. %Multiple coherent round-trips within the cavity amplify the intra-cavity fields, allowing for high photon-pair generation rates even with moderate pump power. The small mode volume enhances the nonlinear interaction, leading to brighter photon sources \cite{bogaerts2012silicon}.
Silicon nitride microresonators are widely employed in nonlinear and quantum optics due to their extremely low losses, strong mode confinement, and moderately high third-order nonlinearity.
Recent demonstrations include the generation of 8 dB of on-chip single-mode quadrature squeezing in a nanophotonic molecule \cite{zhang2021squeezed} and 7.8 dB in a side-wall corrugated resonator \cite{ulanov2025quadrature}, 3.5 dB of intensity-difference squeezing at the detectors (10.5 inferred on chip) using a single pump \cite{shen2025strong} and operating above threshold of parametric oscillation, and 5.6 dB at the detectors below threshold \cite{shen2025highly}.   
Remarkably, their capability to achieve strong quadrature squeezing in a near-single spectral-temporal mode has recently promoted them as the fundamental building block for the generation of Gottesman-Kitaev-Preskill qubits \cite{larsen2025integrated}, forming the resource states for modular continuos-variable (CV) quantum computation \cite{aghaee2025scaling}.
In parallel, the study and characterization of entanglement structure of multiple frequency-modes in squeezed microcombs is rapidly developing. Initially focused on determining correlations between pair of modes \cite{jahanbozorgi2023generation}, the research is now oriented on characterizing the multi-mode structure developing as a consequence of the complex interplay between SFWM and Bragg scattering four-wave mixing  \cite{jia2025continuous,gouzien2023hidden}, as well as to engineer these correlations to build continuous variables cluster states in the frequency domain \cite{wang2025large}.    
All this work has triggered the development of quantum-optical models of microresonators that go beyond single pair generation \cite{BEYOND}. In the high-gain regime, the self-phase modulation of the pump beam (SPM), the cross-phase modulation (XPM) on the squeezed modes, and operator time ordering, play a significant role and their impact can not be neglected. %While these effects have been extensively studied in waveguides, where they notably influence both the mean photon number and temporal mode structure []. However, this work has largely focused on continuous-wave (CW) pumping, leaving the impact of time-dependent effects on temporal mode structure relatively unexplored.\\
While high-gain effects in microresonators have been theoretically \cite{BEYOND,sloan2025high,vendromin2024highly,Strongly_Zachary,vernon2019scalable} and 
experimentally \cite{guidry2022quantum,ramelow2019strong} investigated in the continuous wave (CW) regime, only a few theoretical works have addressed the problem under pulsed excitation \cite{kim2025simulating,cui2021high}, and a systematic experimental study is still lacking. However, pulsed pumping is of great relevance to ensure single temporal mode emission and time-synchronization among arrays of sources, which are necessary features for high-fidelity resource state preparation in scalable CV quantum computing \cite{aghaee2025scaling}, including gaussian boson sampling \cite{arrazola2021quantum}.\\
In this work we experimentally investigate high-gain effects in the generation of pulsed two-mode squeezed light from a silicon nitride microresonator, achieving up to 16 average photons per pulse at the output waveguide, and a maximum inferred on-chip squeezing level of $5$ dB, this is pretty much the limit we could achieve with an escape efficiency of 0.75. We characterized the average number of photons generated per pulse, the second- and first-order self-correlation of each squeezed mode with increasing pulse energy, across various frequency detunings and pulse durations. We found many distinctive signs of the high-gain regime - which were theoretically predicted in the both the CW \cite{Strongly_Zachary} and pulsed \cite{kim2025simulating, f2y1-scgw} excitation - including the observation of a local maximum in the signal/idler generation rate as a function of input pump power, high spectral purity with  detuned pump excitation, and the emergence of a broad pedestal in the first-order coherence function, corresponding to SPM/XPM induced spectral-splitting.
The non-energy-conserving peaks in the joint spectral intensity (JSI), predicted in the high-gain regime  as a result of multi-pair generation in frequency-resolved coincidence measurements \cite{Strongly_Zachary}, appear in our time-resolved experiment as a background of signal and idler photons arriving at the detectors with reduced time correlations compared to those predicted by the joint temporal intensity (JTI). 
We show that by measuring the marginal probability distributions of time-resolved multi-photon events, one can subtract this background contribution and recover to some approximation the time-correlations predicted by the JTI.

%and the joint temporal intensity. We found that SPM and XPM lead to a saturation of the generation rate at approximately $\sim$ 4 average photons/pulse and increase the number of temporal modes. However, for each pump power, there exists an optimal pump frequency detuning relative to the cold cavity resonance that maximizes the generation rate. Notably, at the optimal detuning, the number of temporal modes is also reduced. 

\section{Generation of pulsed squeezed light in lossy microresonators}
\label{sec:theory}
\begin{figure}[]
    \includegraphics[width=\linewidth]{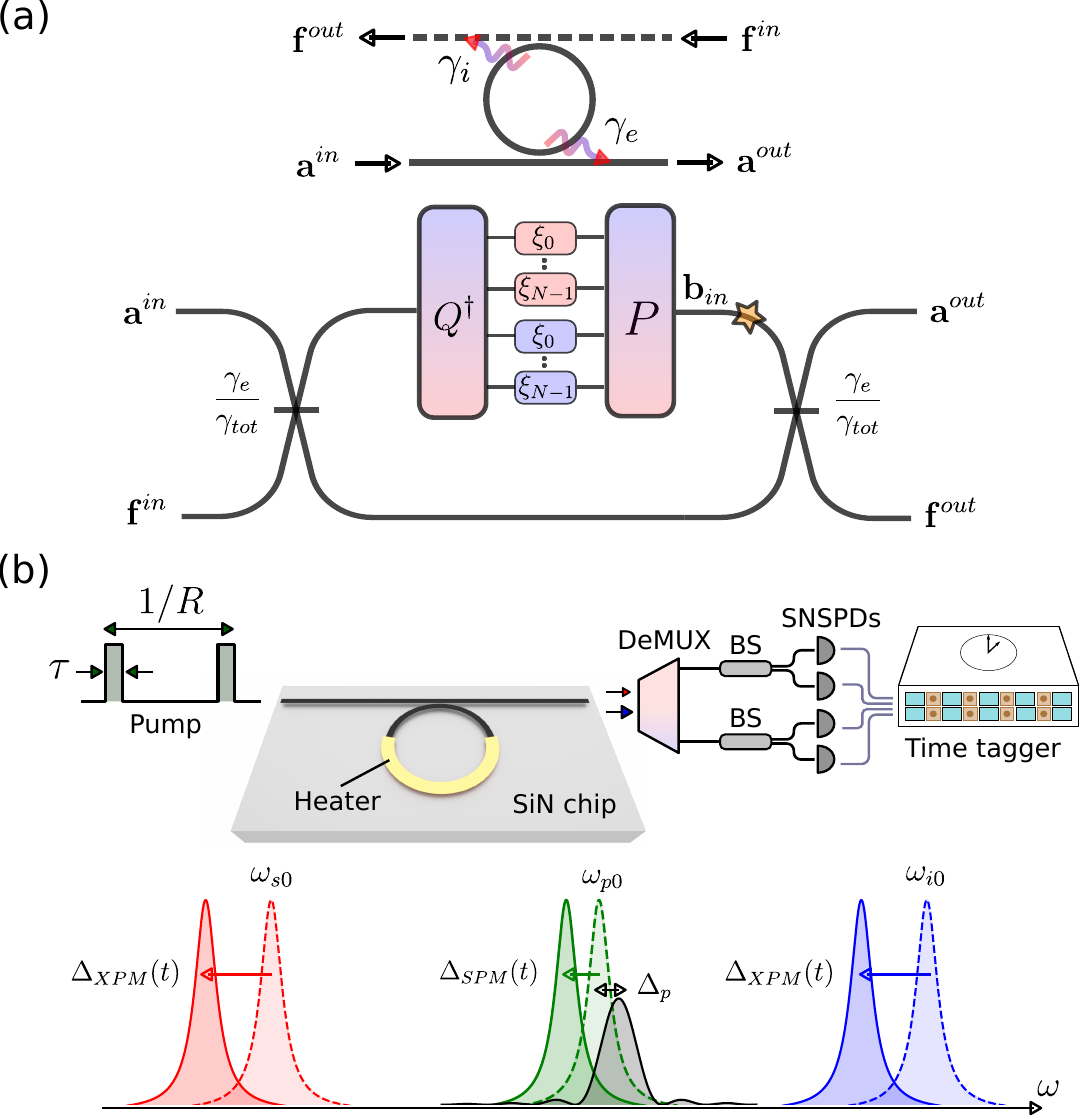}
    \caption{(a) Equivalent circuit representation of the sequence of operations transforming the input operators $(\mathbf{a}^{in},\mathbf{f}^{in})$ to the output operators $(\mathbf{a}^{out},\mathbf{f}^{out})$. (b) Top panel: sketch of the device layout and of the experimental setup. DeMUX: demultiplexer, BS: beasmplitter, SNSPD: superconducting nanowire single photon detector. Bottom panel: sketch of the pump (green), signal (red) and idler(blue) nonlinear resonances shifts $\Delta_{SPM}(t)$ and $\Delta_{XPM}(t)$ induced by SPM and XPM. The cold resonance frequencies are $\omega_{p0}$, $\omega_{s0}$ and $\omega_{i0}$ respectively. The pump spectrum, shown in gray, has the maximum detuned by $\Delta_p$ with respect to the cold cavity resonance frequency $\omega_{p0}$.     
    }
    \label{Fig_0}
\end{figure}
The system under study is a lossy high-Q microresonator coupled to a bus waveguide, as pictured in Fig.\ref{Fig_0}(a). The energy coupling rate into the bus waveguide is $\gamma_e$, while material and scattering losses are modeled by an equivalent \emph{phantom} channel coupled to the resonator with a rate $\gamma_i$ \cite{banic2022two}. When the resonator is resonantly excited through the bus waveguide with a bright coherent state at central frequency $\omega_p$, photon pairs are generated by SFWM, occupying the many resonance frequencies that simultaneously satisfy energy and momentum conservation (we use the standard convention of calling the photon with a lower wavelength idler, and the one with a higher wavelength signal). In what follows, we will focus on a single pair of resonances at frequencies $\omega_{s0}$ and $\omega_{i0}$. In general, if the system is pumped below the threshold of optical parametric oscillation (OPO), a multimode squeezed state is generated within each resonance, which through the Bloch-Messiah (BM) decomposition can be decomposed into the tensor product of many independent two-mode squeezers (TMS), also called the squeezed supermodes of the system \cite{cui2021high}. We remark that these supermodes refer to a BM decomposition within a single pair of resonances, thus differing from the definitions in \cite{jia2025continuous} or \cite{gouzien2023hidden}, where they span multiple resonances. The effect of loss can be equivalently modeled by letting the signal/idler supermodes to impinge on a beasmplitter with transmittivity  $p_e=\frac{\gamma_e}{\gamma_{tot}}$  (where $p_e$ is the escape efficiency and $\gamma_{tot}=\gamma_e+\gamma_i$), and by tracing out the reflected modes \cite{cui2021high,Brusaschi:24}. The system dynamics involving the input(output) waveguide field operators $\textbf{a}^{in(out)}$, the cavity field operators $\mathbf{c}$ and the input/output phantom channel field operators $\mathbf{f}^{in(out)}$ has been solved both in the frequency \cite{kim2025simulating,cui2021high} and in the time domain \cite{BEYOND,Strongly_Zachary}. In particular, the frequency picture explicitly highlights the relation between the set of input operators $(\mathbf{a}^{in},\mathbf{f}^{in})$ and the output operators $(\mathbf{a}^{out},\mathbf{f}^{out})$, which is given by the a symplectic transformation $(\mathbf{a}^{out},\mathbf{f}^{out})^T= \mathbf{S}(\mathbf{a}^{in},\mathbf{f}^{in})^T$ (here we used the bold symbol, such as $\mathbf{a}$, as a short-hand for the row vector of annihilation and creation operators $[a_{s,0},...,a_{s,N-1},a_{i,0}^{\dagger},...,a_{i,N-1}^{\dagger}]$, where s(i) labels the signal(idler) photon and the numbers from $0$ to $N-1$ indicate the frequency, which is discretized in steps of $\Delta\omega$ as $\omega_{s(i),k}=\lceil \omega_{s(i)}\rceil+k\Delta\omega$) ($\lceil \omega_{s(i)}\rceil$ is the lower bound of the signal(idler) frequency interval) \cite{kim2025simulating}. The symplectic matrix $\mathbf{S}$ can be written as $\mathbf{S}=\mathbf{U_2CU_1}$, where $\mathbf{U}_{1(2)}$ describes passive beamsmplitter transformations with transmittivity $\gamma_e/\gamma_{tot} $ ($\mathbf{U}_2=-\mathbf{U}_1$ in the high-Q limit) and $\mathbf{C}$ describes the transformation applied by a lossless twin-beam parametric amplifier  \cite{kim2025simulating,cui2021high}. The matrix $\mathbf{C}$ can be factored through the BM decomposition as $\mathbf{C}=\mathbf{Q}^{\dagger}\mathbf{R}\mathbf{P}$, where $\mathbf{Q}$ and $\mathbf{P}$ describe passive unitary transformations, and
\begin{equation}
    \mathbf{R} = 
    \begin{pmatrix}
\textrm{cosh}(\bm{\xi}) & \textrm{sinh}(\bm{\xi}) \\
\textrm{sinh}(\bm{\xi}) & \textrm{cosh}(\bm{\xi})
\end{pmatrix}, \label{eq:matrix_R}
\end{equation}
with $\bm{\xi}=\textrm{diag}(\xi_{0},...,\xi_{N-1})$ contains the $N$ squeezing parameters $\xi_{i=0}^{N-1}$. This series of transformations can be represented by the optical circuit shown in Fig.\ref{Fig_0}(a). \\
The key result is that, below the threshold of parametric oscillation, the dynamics are described by a series of gaussian operations acting on the vacuum; hence, the output state is still gaussian. In particular, these relations incorporate nonlinear resonance shifts arising from self- and cross-phase-modulation (SPM/XPM), naturally handles time-ordering corrections, and relies only on the undepleted pump approximation \cite{kim2025simulating}.   
The columns of $\mathbf{P}$ then define the supermodes, representing the set of Schmidt modes of the overall system, which characterize the spectral (temporal) mode structure of the resonator. \\
In our work, we will focus on metrics that are more naturally expressed in the time domain rather than in the frequency domain, such as the time-dependent average photon number or the joint temporal intensity. For this reason, we solve the system dynamics by numerically integrating the master equation for the signal/idler density matrix in the time domain, following the approach outlined in \cite{Brusaschi:24}.\\
In the interaction picture, the Hamiltonian describing the process of XPM and SFWM on the signal/idler cavity operators $c_s$ and $c_i$ is given by \cite{BEYOND,Brusaschi:24}
\begin{equation} \label{H_non_lineare}
\begin{split}
    H_{\textup{NL}}/\hbar ={} & -[\Lambda (\langle c_p (t) \rangle^{*} )^2 e^{i\Delta\omega_D t}c_s^{\dagger}c_i^{\dagger} +\textrm{h.c.}] \\
    & - \Delta_{XPM}(t) (c_s^{\dagger} c_s + c_i^{\dagger} c_i),
\end{split}
\end{equation}
where $\Delta_{XPM}=2\Lambda|\langle c_p(t) \rangle|^2$, $\Delta\omega_D=4\pi D_{\textrm{int}}m^2$  ($D_{\textrm{int}}=-1.38\times10^{-6}\,\textrm{ps}^{-1}$ is the integrated dispersion \cite{BEYOND}, calculated from Finite Element Method simulations, and $m=5$ is the number of free spectral ranges (FSR) from the pump resonance of the signal/idler beams), and
\begin{equation}
    \Lambda = \frac{\hbar \omega_p^2 c n_2}{n_0^2 V_{\textup{eff}}},
\end{equation}
describe the nonlinear coupling rate associated with SFWM, where $n_2$ is the nonlinear refractive index of the material, $V_{\textup{eff}}$ is the effective mode volume of the optical mode in the resonator and $n_0$ the refractive index.
The C-number $\langle c_p (t) \rangle  $ is the function describing the average number pump photons inside the resonator and is obtained by numerically integrating the differential equation
\begin{equation}
    \bigg [\frac{d}{dt} + \frac{\gamma_{tot}}{2} -i\Delta_{SPM}(t) \bigg ] \langle c_p(t) \rangle = -i \sqrt{2 \gamma_{e}} \beta (t) e^{i \Delta_p t}. \label{eq:pump_equation}
\end{equation}
The term $\Delta_{SPM}(t)=\Lambda|\langle c_p(t) \rangle|^2$ accounts for SPM on the pump beam, $\Delta_p=\omega_p-\omega_{p0}$ is the relative detuning of the central pump frequency $\omega_p$ to the cold resonance frequency $\omega_{p0}$, and $\beta(t)$ describes the driving field.
%To write Eq. \ref{H_non_lineare} we replaced the pump operators with the complex numbers, $\langle a_p \rangle \xrightarrow{} a_p$ and $\langle a_p \rangle^* \xrightarrow{} a_p^{\dagger}$, being a strong coherent state. 
Pump depletion terms in Eqs.(\ref{eq:pump_equation},\ref{H_non_lineare}) are neglected because we will focus on regimes where the mean number of signal/idler photons in the cavity is much smaller than the number of pump photons. %By ignoring this term we cannot correctly predict when the cavity enters in the Optical Parametric Oscillator (OPO) regime. We are not interested in studying that regime so we work experimentally below the OPO threshold.\\
We consider as driving field $\beta (t)$ a top-hat function of duration $T$, which approximates the rectangular pulses used in the experiments and expressed as
\begin{equation}
    \beta (t) = \sqrt{\frac{P}{R \hbar \omega_p }} (\Theta (t)- \Theta (t- T)), \label{eq:beta_pump}
\end{equation}
where $P$ is the average power, $R$ is the repetition rate, and $\Theta$ is the Heaviside step function. \\
The master equation for the signal and idler cavity modes $c_{s(i)}$ in the Lindblad form is \cite{Brusaschi:24}
\begin{equation}
    \frac{d \rho}{dt} = -\frac{i}{\hbar} [H_{\textup{NL
    }}, \rho] + (\mathcal{D}[\sqrt{\gamma_{tot}} c_s] + \mathcal{D}[\sqrt{ \gamma_{tot}} c_i]) \rho,  \label{eq:master_equation}  
\end{equation}
where $\mathcal{D}[c]=c\rho c^{\dagger}-\frac{1}{2}\{c^{\dagger}c,\rho\}$ and $\{\cdot,\cdot\}$ denotes and anti-commutator. The master equation is numerically integrated using the open access Python library \texttt{QuTiP} \cite{QuTiP}.
Any simulated observables and metric discussed in the next sections are calculated from the density matrix.
The output operators $a^{out}(t)$ in Fig.\ref{Fig_0}(a) are related to the cavity operators $c(t)$ from the standard input-output relation $a^{out}(t)=a^{in}(t)+i\sqrt{2\gamma_e}c(t)$.

\section{\label{sec:experimental_res} Experimental results}
%In this section, we show the experimental results and compare them to numerical simulations done following the approach from the previous section.
%We measured almost identical linear responses for the pump, signal, and idler modes within the ring resonator
The silicon nitride microresonator under study is realized by a waveguide with cross section $1.8\times0.8\,\mu m^2$, ($\Lambda\sim1.4$ Hz) having slightly anomalous dispersion at the pump wavelength. The resonator has a FSR of 200 GHz and is overcoupled to the bus waveguide, with an escape efficiency of $p_{e} = 0.75$ at the pump, signal and idler modes. We measured a loaded quality factor of $Q=8\times10^5$ ($1/\gamma_{tot}\sim660$ ps), and an intrinsic quality factor of about $3\times10^6$ ($1/\gamma_i\sim2730$ ps). We choose to work with a pump wavelength of $\lambda_p = 1544.53$ nm, while the squeezed modes under consideration lie at the signal and idler wavelengths of $\lambda_i = 1536.50$ and $\lambda_s=1552.50$.  %For the simulations, we assumed non-linear coefficients $\gamma_{sfwm}$, $\gamma_{spm}$,$\gamma_{xpm,i}$,$\gamma_{xpm,s}$
%equal to 1 $ W^{-1} m{-1}$.
The resonator is pumped by strong rectangular optical pulses of time duration $T$, and with a fixed repetition rate of $R=100$ kHz. The pulses are carved from a continuous wave butterfly laser diode using an amplitude electro-optic modulator (EOM), and subsequently amplified by an Erbium-Doped Fiber Amplifier. 
We choose to work with a low repetition rate to reduce the average power (on the order of $100\,\mu W$ in most of the experiments), thus mitigating the average resonance shift due to the thermo-optic effect. Moreover, the short pulse duration ensures that thermal effects can be neglected also within each pump pulse, as the characteristic thermal response time of the resonator (tipically in the order of hundrends of ns for SiN resonators of similar size \cite{stone2025reduction}) is much smaller than the pulse duration. 
After removal of the background noise at the signal/idler wavelengths, the pump power is regulated by an electronic variable optical attenuator (VOA) and coupled to the chip using Ultra High Numerical Aperture (UHNA4) fibers and inverse tapers, achieving a coupling loss of $\sim 1$ dB/facet. Signal and idler photons generated by single-pump SFWM are separated by passband filters, which also suppress the residual pump light, and are finally detected using superconducting nanowire single-photon detectors (SNSPDs). A time-tagger correlator is used to register their arrival time relative to the pump trigger.
%Thermal effects could, in principle, contribute to shifting the resonance wavelength, making it difficult to understand which phenomenon we are observing. 

\subsection{Time-dependent average photon number}
\begin{figure*}[t!]
    \includegraphics[width=\textwidth]{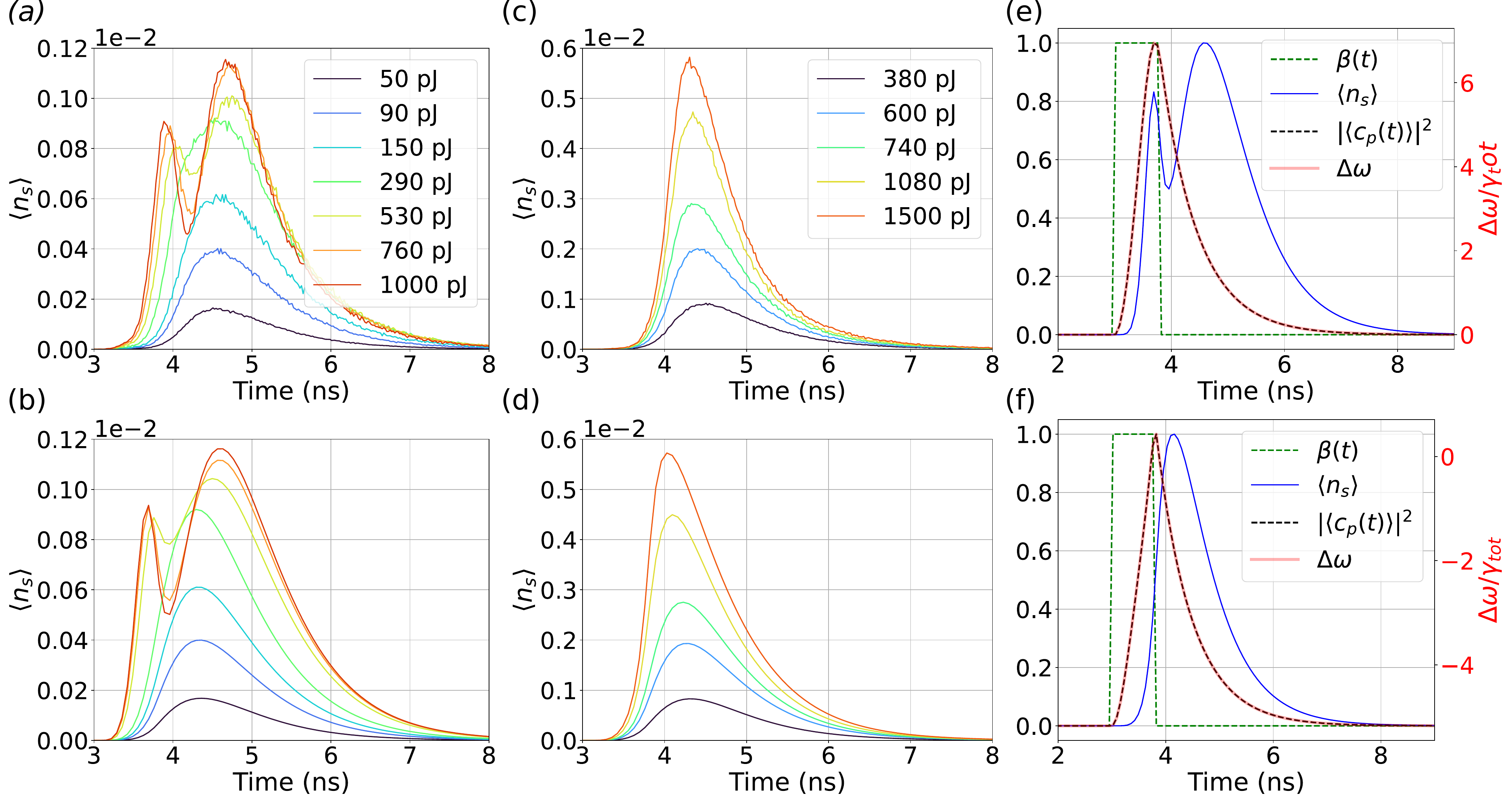}
    \caption{(a) Measured average number of photons $\langle n_s (t)\rangle$ in the marginal signal beam as a function of time for different pump energies (both quantities are estimated on-chip) and $\Delta_p=0$. (b) Calculated $\langle n_s (t)\rangle$ at the same pulse energies indicated in panel (a) and with $\Delta_p=0$.
    (c) Measured and calculated (panel (d)) value of $\langle n_s (t)\rangle$ as a function of time for different pump energies at $\Delta_p=\Delta_{opt}$. (e) Simulation showing the incident pump envelope $\beta (t)$, the number of pump photons in the cavity $|\langle c_p(t)\rangle|^2$, the average number of signal photons $\langle n_s(t)\rangle$ in the output waveguide and the the energy mismatch $\Delta\omega$ (referred to the right vertical axis) for $\epsilon_p=1000$ pJ and $\Delta_p=0$. The values of $\beta (t)$, $|\langle c_p(t)\rangle|^2$ and $\langle n_s(t)\rangle$ have been all normalized to their maximum value for improved visualization. The same quantities are shown in panel (f) for $\epsilon_p=1500$ pJ and $\Delta_p=\Delta_{opt}.$
    }
    \label{Fig_1}
\end{figure*}
\begin{figure}[]    \includegraphics[width=\columnwidth]{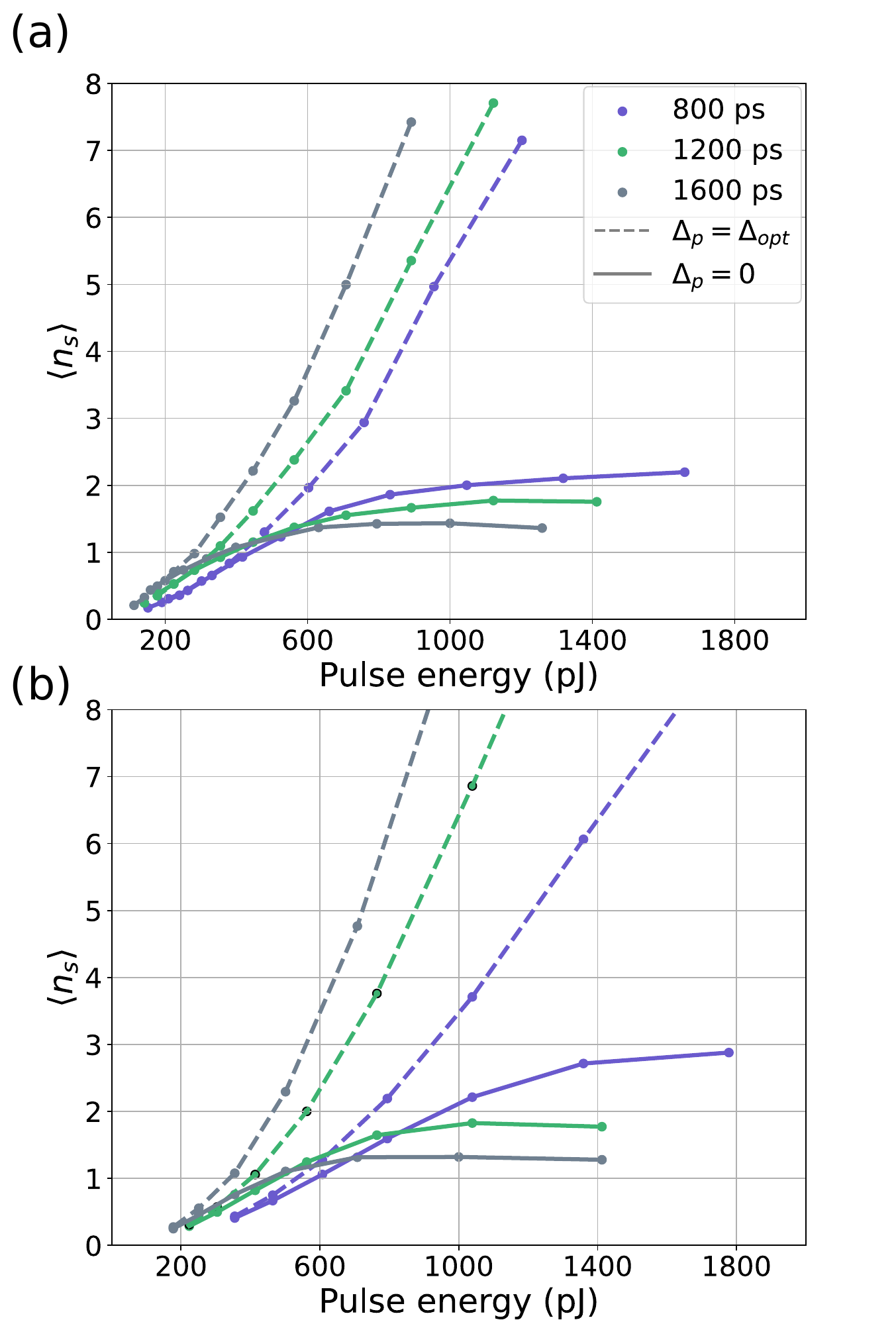}
    \caption{(a) Measured average number of signal photons per pulse in the output waveguide as a function of the input pulse energy for different pulse durations. The dashed lines indicate the measurement performed at $\Delta_p=\Delta_{opt}$, while the continuous line is the measurement at $\Delta_p=0$. Error bars are smaller than the size of the data symbols. (b) Simulations of the average number of photons per pulse as a function of the input pulse energy for different pulse durations. Colors and line-style have the same meaning as in panel (a).
    }
    \label{Fig_2}
\end{figure}
\begin{figure}[]
    \includegraphics[width=\columnwidth]{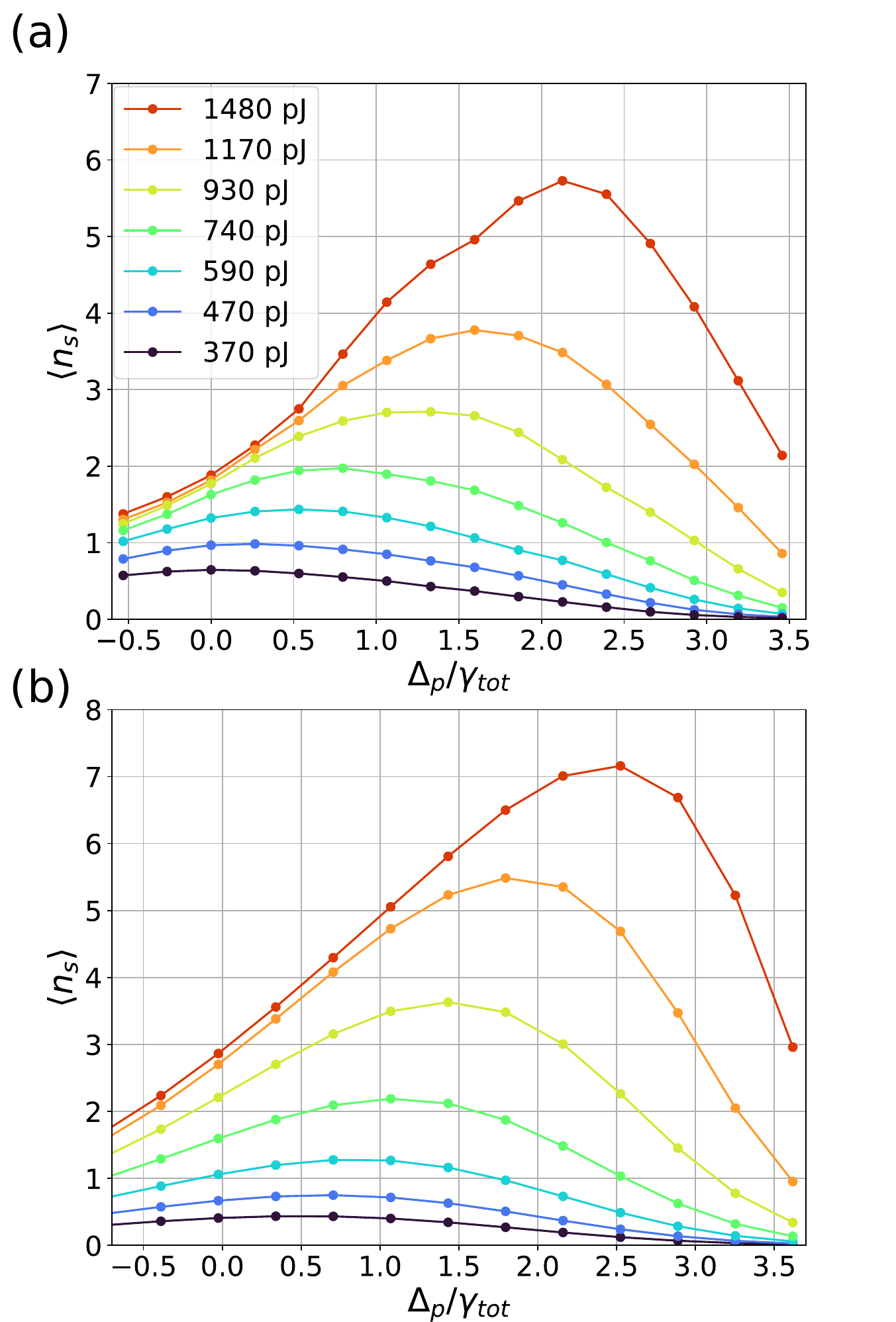}
    \caption{(a) Measured average number of signal photons in the output waveguide as a function of the pump detuning $\Delta_p$ for different input powers. Error bars are smaller than the size of the data symbols. (b) Simulated average number of signal photons in the output waveguide as a function of the pump detuning $\Delta_p$.
    }
    \label{Fig_3}
\end{figure}
The first quantity that we characterize is the average number of photons in one of the marginal beams at the output waveguide and their time-dependence at different pump pulse energies, pump detunings $\Delta_p$ and pulse duration $T$. Without loss of generality, we will focus on the signal beam. At first, the pulse duration is set to $T=800$ ps, i.e. comparable to the photon dwelling time in the resonator, and the detuning to $\Delta_p=0$, which is the value that maximizes the pair generation rate at low parametric gain.  %In this paper we investigate the average number of photons generated and its time evolution in different regimes. We focus mainly on studying how detuning the pump laser used to generate the state from the resonance wavelength leads to interesting behavior of different observables and different effects.\\
The input pulse excites the resonator at $t\sim3$ ns, corresponding to the initial increase of the signal intensity.
In Fig.\ref{Fig_1}(a,b) we report the measured and calculated time-dependent average number of photons per pulse $\langle n_s (t) \rangle$ for different pump pulse energies $\epsilon_p$. For $\epsilon_p<300$ pJ, the generation rate increases over time up to a maximum, and then exponentially decreases, following the characteristic decay time of the cavity. However, for $\epsilon_p>300$ pJ, the shape of $\langle n_s(t)\rangle$ changes and becomes bimodal, with the emergence of a narrow peak at short times followed by a broader and more intense peak at later times, and whose separation increases with increasing pulse energy. For $\epsilon>760$ pJ, the generation rate saturates and the time-dependence no longer changes. \\
It is found that for each pulse energy there is an optimal value of the pump frequency detuning $\Delta_p=\Delta_{opt}$ that maximizes the generation rate, in accordance with the theoretical predictions in both the continuous wave \cite{Strongly_Zachary} and the pulsed regime \cite{kim2025simulating}. The pump detuning compensates the SPM and XPM resonance shifts introduced by the high circulating power in the resonator. The behavior of $\langle n_s(t) \rangle$ for $\Delta=\Delta_{opt}$ is characterized as a function of $\epsilon$. The results, reported in Fig.\ref{Fig_1}(c), show that the temporal profile of $\langle n_s(t)\rangle$ does not change with increasing pulse energy (all the curves are simply scaled), in sharp contrast with the behavior at $\Delta_p=0$. To explain the different dynamics, we simulated $\langle n_s(t)\rangle$ for both $\Delta_p=0$ and $\Delta_p=\Delta_{opt}$ by numerically integrating Eq.(\ref{eq:master_equation}), and used the density matrix to calculate $\langle n_s (t)\rangle=\textrm{Tr}(\rho(0)a^{out}_s(t)^{\dagger}a^{out}_s(t))$. The results of the simulation are shown in Fig.\ref{Fig_1}(b) for $\Delta_p=0$ and $\epsilon=1000$ pJ, and in Fig.\ref{Fig_1}(d) for $\Delta_p=\Delta_{opt}$ and $\epsilon=1500$ pJ. Moreover, to highlight the impact of SPM and XPM, we also reported in Fig.\ref{Fig_1}(e,f) the incident pump envelope $\beta(t)$, the number of pump photons in the cavity $|\langle c_p(t)\rangle|^2$, and the energy mismatch $\Delta\omega=2(\omega_p(t)-\Delta_p)-\omega_s(t)-\omega_i(t)$, where $\omega_{s(i)}(t)=\omega_{s(i)0}-\Delta_{XPM}(t)$ and $\omega_p(t)=\omega_{p0}-\Delta_{SPM}(t)$ (see Fig.\ref{Fig_0}(b)). The energy mismatch determines the efficiency of the SFWM process \cite{BEYOND}, which is maximum for $\Delta\omega=0$. The behavior of $|\langle c_p(t)\rangle|^2$ is quite similar for both $\Delta_p=0$ and $\Delta_p=\Delta_{opt}$: the internal pump energy builds up until the end of the incident pump pulse, and then it decreases exponentially following the characteristic decay time of the cavity $1/\gamma_{tot}$. What discriminates the behavior of $\langle n_s(t)\rangle$ for $\Delta_p=0$ and $\Delta_p=\Delta_{opt}$ is $\Delta\omega$. 
When $\Delta_p=0$ (Fig.\ref{Fig_1}(e)), the energy mismatch increases with increasing internal pump energy because of the SPM/XPM induced resonance shifts, thus it is maximized when $|\langle c_p(t)\rangle|^2$ reaches its highest value. In this regime, the efficiency of SFWM is governed by a trade-off between minimizing energy mismatch and maximizing pump intensity. The optimal balance occurs at two distinct moments: slightly before the end of the pump pulse, where the internal energy is close to its maximum value but the energy mismatch is large ($\Delta\omega/\gamma_{tot}\sim6.5$), and during the decay of the internal energy, where $\Delta\omega/\gamma_{tot}\sim1.25$. This results in the double peak structure in $\langle n_s (t)\rangle$ observed in Fig.\ref{Fig_1}(a), and well reproduced in the simulations reported in Fig.\ref{Fig_1}(b). 
%in the temporal behavior of the average number of photons generated. In the paper, we will refer to this pump regime in which the wavelength of the pump matches perfectly the wavelength of the cold resonance as the zero detuning regime.\\
In contrast, when $\Delta_p=\Delta_{opt}$, the simulation reported in Fig. \ref{Fig_1}(f) shows that the energy mismatch is minimum when the field intensity is nearly maximized. Therefore, the initial detuning compensates the resonance shift induced by SPM/XPM, resulting in a single peak structure in the average number of photons in both the experiment (Fig.\ref{Fig_1}(c)) and simulation (Fig.\ref{Fig_1}(d)).\\ %and also a condition in which the average number of photons over the full time pulse is maximized. 
We then characterized the average number of signal photons $\langle n_s  \rangle$ per pulse, which corresponds to the area under the curves of $\langle n_s(t)\rangle$ in Fig.\ref{Fig_1}.
%\begin{equation} \label{nmed}
%    \langle n_s  \rangle = \int_0^{+\infty} dt \langle a_s^+(t) a_s(t) \rangle
%\end{equation}
%We can simulate, starting from the density matrix that we calculate with QuTiP, the average number of scattered photons in the channels waveguide per pulse integrated over time $\langle n_s^{simulated} \rangle$ as
%\begin{equation}
%   \langle n_s^{simulated} \rangle = 2 \gamma_{es} \int Tr(\rho(t) a_s^{\dagger } a_s ) dt
%\end{equation}
The characterization is performed at different pulse energies, for the pulse durations $800$ ps, $1200$ ps and $1600$ ps, and by setting the pump detuning to either $\Delta_p=0$ or $\Delta_p=\Delta_{opt}$.
The experimental results, shown in Fig.\ref{Fig_2}(a), indicate that for $\Delta_p=0$ the average number of signal photons saturates around $2-3$ photons per pulse for all the investigated pulse durations. This is explained by the increasing energy mismatch $\Delta\omega$ induced by SPM and XPM, which develops as the pulse energy increases. The saturation of $\langle n_s\rangle$ is in agreement with the theoretical predictions in \cite{kim2025simulating}. \\
In contrast, for $\Delta_p=\Delta_{opt}$, we observe that $\langle n_s  \rangle$ monotonically grows as function of the input pulse energy. The endpoints of the curves in Fig.\ref{Fig_2}(a) for $\Delta_p=\Delta_{opt}$ lie slightly below the power threshold of optical parametric oscillation (which is seen to decrease with the pulse duration up to $T\sim10$ ns, where it reaches its CW value of $\sim35$ mW). %When crossing this threshold, we found that even a $3\%$ increase of the pump power leads to an increase of $\langle n_s  \rangle$ by more than one order of magnitude. 
Below threshold, $\langle n_s  \rangle$ grows as $\epsilon^2$, in agreement with the theoretical predictions in both the CW \cite{Strongly_Zachary} and pulsed regime \cite{kim2025simulating}. Interestingly, while for $\Delta_p = 0$, shorter pulses lead to a saturation of $\langle n_s \rangle$ at higher values, the opposite behavior is observed for $\Delta_p = \Delta_{\text{opt}}$, where, for a fixed pulse energy, $\langle n_s \rangle$ increases with increasing pulse duration. Thus, working with pulses that are twice as long as the cavity lifetime could be beneficial for applications where high-photon fluxes are required.  The simulations, reported in Fig.\ref{Fig_2}(b), align very well with the experimental results.\\
Finally, we characterized the average number of photons as a function of the pump detuning $\Delta_p$ for $T=800$ ps. In Fig.\ref{Fig_3}(a) we report the measured data, while in Fig.\ref{Fig_3}(b) the simulations. As expected, the optimal detuning for which $\langle n_s\rangle$ is maximized monotonically increases with the pulse energy. The relation between $\Delta_{opt}$ and $\epsilon_p$ is linear, which is expected since $\Delta\omega \propto |\langle c_p\rangle|^2$ and $|\langle c_p\rangle|^2\propto\epsilon_p$ at the detuning which maximizes the generation rate \cite{Strongly_Zachary}.
When $\langle n_s\rangle\sim0.5$,  $\Delta_{opt}\sim\gamma_{tot}/2$, which can be identified as a boundary where SPM and XPM effects can not be neglected, while for $\langle n_s\rangle\ge2$ the optimal detuning becomes larger then the linewidth of the pump resonance. 
%When the resonance is pumped with stronger intensity, the XPM and SPM effects become more impacting, indeed we observe that the detuning needed to maximize the generation is higher. At low input power by matching the cold resonance wavelength perfectly maximizes the generation, but this is not true anymore when the average number of photons is over 0.5. This is in agreement with the measurements and the simulation in Fig. \ref{Fig_1}.\\
\subsection{Second-order correlation of the marginal beams}
\begin{figure}[]    \includegraphics[width=\columnwidth]{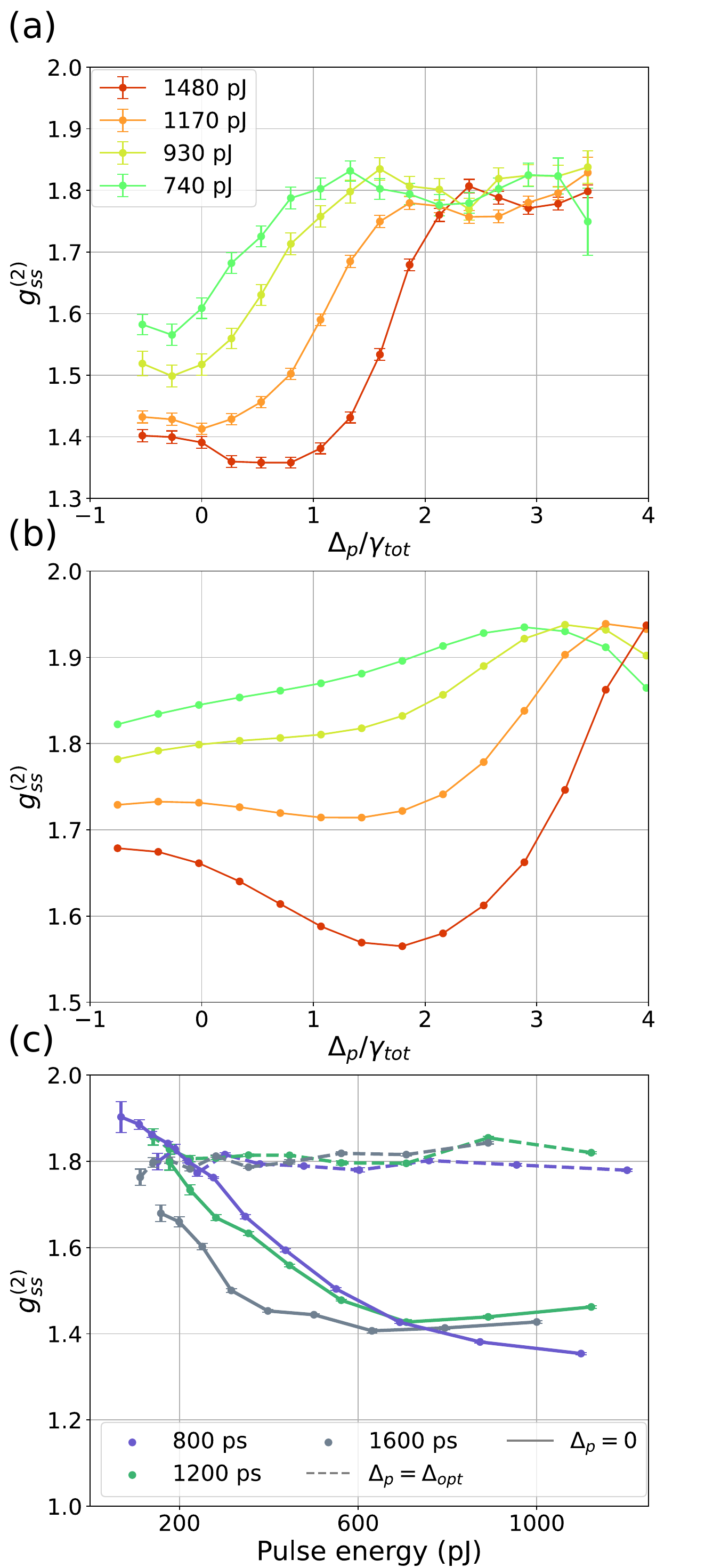}
    \caption{(a) Measured unheralded second-order correlation $g_{ss}^{(2)}$ of the signal  mode as a function of the detuning between the pump and the cold resonance frequency for different pump pulse energies. (b) Numerical simulations of $g_{ss}^{(2)}$ for the same pump pulse energies shown in panel (a). (c)  Measured unheralded second-order correlation of the $g_{ss}^{(2)}$ signal mode for different pulse durations as a function of the input pulse energy. The dashed lines report the measurements for $\Delta_p=\Delta_{opt}$, while the continuous lines for $\Delta_p=0$.
    }
    \label{Fig_4}
\end{figure}
The second-order correlation correlation of either the signal or the idler mode is  defined as
\begin{equation}
    g_{ss}^{(2)} = \frac{\langle n_s^2  \rangle -\langle n_s  \rangle}{\langle n_s  \rangle^2} \label{eq:g2_definition}
\end{equation}
This quantity, frequently called the unheralded second-order correlation, is related to the amplitudes $\xi$ of the squeezed modes as \cite{christ2011probing,kim2025simulating}
\begin{equation} \label{gdos}
    g_{ss}^{(2)} = 1 + \frac{\sum_k \textrm{sinh}^4(\xi_k)}{\left(\sum_k \textrm{sinh}^2(\xi_k)\right)^2}
\end{equation}
The second term on the right hand side of Eq.(\ref{gdos}) is called the Schmidt number $\mathcal{K}$, and quantifies the effective number of squeezed modes. The Schmidt number is related to the spectral purity $\mathcal{P}$ of the marginal state as $\mathcal{K}=1/\mathcal{P}$, which allows one to write $g_{ss}^{(2)}= 1+\mathcal{P}$.
One can expand the operator $n_s$ in Eq.(\ref{eq:g2_definition}) in terms of $a_s^{out}(t)$ and $a_s^{out}(t)^{\dagger}$ to obtain \cite{christ2011probing,laiho2022measuring}
\begin{equation}
    g_{ss}^{(2)} = \frac{\int dtd\tau \langle a_s^{out}(t)^{\dagger} a_s^{out}(t+\tau)^{\dagger} a_s^{out}(t+\tau) a_s^{out} (t) \rangle}{(\int dt \langle a_s^{out}(t)^{\dagger} a_s^{out}(t) \rangle)^2},  
\end{equation}
which relates $g^{(2)}_{ss}$ to single and coincidence count probabilities $p_{\textrm{single}}$ and $p_{\textrm{coinc}}$ (in the limit where \mbox{$p_{\textrm{single}(\textrm{coinc})}\ll 1$)} as $g^{(2)}_{ss}=p_{\textrm{coinc}}/p_{\textrm{single}}^2$ \cite{laiho2022measuring}.
We measured the unheralded second-order correlation function \( g_{ss}^{(2)} \) using the setup shown in Fig.\ref{Fig_0}(b). This was done by recording the click probability between the two outputs of the 50/50 beamsplitter to which the signal photons were demultiplexed. The coincidence probability was then normalized by the product of the photodetection probability of each individual output.
Similarly to what was done for $\langle n_s\rangle$, the values of \( g_{ss}^{(2)} \) were measured for different input powers and pump detunings, initially setting the pulse duration to $800$ ps. The results of this characterization are reported in Fig.\ref{Fig_4}(a). %We can directly compared with those in Fig. \ref{Fig_3}, as both measurements were taken simultaneously under identical conditions. 
By comparing them to the behavior of $\langle n_s\rangle$ in Fig.\ref{Fig_3}(a), we observe that for each pump power the \( g_{ss}^{(2)} \) (and thus the associated spectral purity $\mathcal{P}$) reaches a maximum approximately at the same pump detuning where the average photon number is maximized. %Then by going to a positive detuning the $g_{ss}^{(2)}$ grows up until it gets to maximum exactly when the average photon number is maximized
In contrast with $\langle n_s\rangle$, which rapidly decreases for $\Delta >\Delta_{opt}$, the $g^{(2)}_{ss}$ remains approximately constant. Moreover, for pump detuning close to $\Delta_{opt}$ the $g^{(2)}_{ss}$ is less sensitive to the pulse energy. Thus, by operating at $\Delta_p = \Delta_{opt}$, one can simultaneously maximizes both the average photon number and the spectral purity, as theoretically predicted in \cite{kim2025simulating}. We supported these results through numerical simulations of \( g_{ss}^{(2)} \), which were obtained by first calculating the JTA, then performing a singular value decomposition to extract the squeezing parameters \( \xi_k \) from the resulting diagonal matrix (see Appendix A for more details), and finally inserting these quantities in Eq.\ref{gdos}.  
The simulations are shown in Fig. \ref{Fig_4}(b). While the qualitative behavior observed in the experiment aligns well with the simulation, the experimental value of the spectral purity in Fig.\ref{Fig_4}(a) is generally lower than the theoretical one. We believe this is due to noise arising from spontaneous Raman scattering, which is known to affect measurments of the second-order correlation \cite{borghi2024uncorrelated}.\\
Finally, Fig.\ref{Fig_4}(c) shows the measured unheralded second-order correlation as a function of the pump pulse energy for different pulse durations, and for $\Delta_p=0$ and $\Delta_p =\Delta_{opt}$. In the first case, the purity decreases with increasing pulse energy for all pulse durations. At low pump energies, for which SPM/XPM shifts are negligible, we recover the well-known result that the spectral purity decreases with increasing the pulse duration \cite{christensen2018engineering}. In contrast, for $\Delta_p=\Delta_{opt}$, the purity remains high across the range of pulse energies, and we do not observe appreciable differences between the different pulse durations. Intuitively this can be explained by observing that at $\Delta_p = 0$, the SPM- and XPM-induced resonance shifts effectively increase the pump pulse duration. For example, one can see from Fig.\ref{Fig_1}(a,c) that the probability of photon emission extends well beyond the time at which the pump energy has decayed to the $1/e\sim0.36$ level. On the contrary, for $\Delta_p=\Delta_{opt}$, the initial pump detuning  compensates the nonlinear resonance shifts, rapidly reducing the photon emission probability after that the pump energy has decayed to the $1/e\sim0.36$ level.
%In conclusion, since for a fixed pump pulse energy and for $\Delta_p=\Delta_{opt}$ the average number of photons increases with the pulse duration (see Fig.\ref{Fig_2}(a)), working with pulses that are twice as long as the cavity lifetime could be beneficial for applications where high-photon fluxes and single-mode emission are required. 

%When a two-mode squeezed (TMS) state is marginalized over one mode, the resulting single-mode state is thermal \cite{loudon2000quantum}. For a TMS state composed of multiple Schmidt modes, the marginal state becomes multithermal, resulting in \( g_{ss}^{(2)} < 2 \). The presence of thermal noise further increases the degree of multithermality, leading to a decrease in the measured \( g_{ss}^{(2)} \).\\

\subsection{First-order correlation of the marginal beams}
\begin{figure*}[]
    \includegraphics[width = \textwidth]{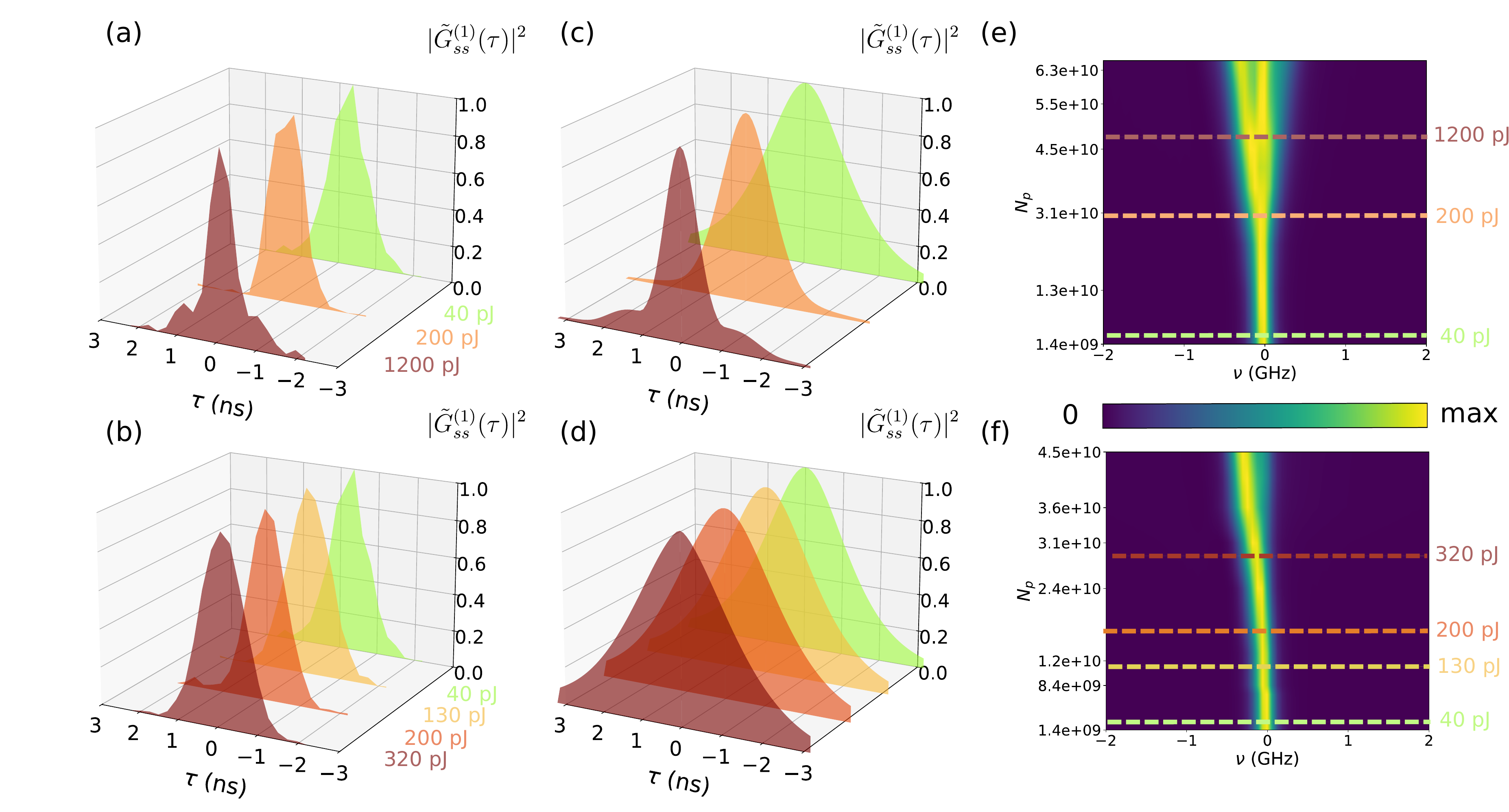}
    \caption{(a) Measured correlation function $\tilde{G}_{ss}^{(1)}(\tau)$ for three different input pump pulse enrgies (40 pJ, 200 pJ and 1200 pJ) and $\Delta_p=0$.  (b) Measured correlation function $\tilde{G}_{ss}^{(1)}(\tau)$ for four different input pump pulse energies (40 pJ, 130 pJ, 200 pJ and 320 pJ) and $\Delta_p=\Delta_{opt}$. 
    (c) Simulated $\tilde{G}_{ss}^{(1)}(\tau)$ using the same input pulse energies shown in panel (a) and $\Delta_p=0$. (d) Simulated $\tilde{G}_{ss}^{(1)}(\tau)$ using the same input pulse energies shown in panel (b) and $\Delta_p=\Delta_{opt}$. (e) Simulated single photon spectrum as a function of the average number of pump photons in the cavity $N_P$ for $\Delta_p=0$ and (f) $\Delta_p=\Delta_{opt}$. The dashed lines indicate the pulse energies investigated in the experiment and shown in Fig.\ref{Fig_5}(a,c). 
    }
    \label{Fig_5}
\end{figure*}
In the high-gain regime, notable modifications in the spectral and related temporal first-order correlations of the emitted photons are expected due to the SPM/XPM and time-ordering effects \cite{Strongly_Zachary,kim2025simulating,BEYOND}. 
Due to the high-Q of the resonator, it was not possible to directly measure the single-photon spectra because of the lack of narrow-band filters with sub GHz resolution. However, spectral modifications can be investigated in the complex (not-normalized) temporal first-order correlation function $G_{ss}^{(1)}(\tau)$, defined as
$G_{ss}^{(1)} (\tau) = \int dt \langle [a_s^{out}(t)]^{\dagger} a_s^{out}(t +\tau ) \rangle$,
by exploiting the fact that the single-photon spectrum and $G_{ss}^{(1)}(\tau)$ are related by a Fourier transform \cite{Strongly_Zachary}. From the characteristic relation \cite{loudon2000quantum}
\begin{equation} \label{da_g2_a_g1}
    \tilde{g}_{ss}^{(2)}(t_1,t_2) = 1 + |g_{ss}^{(1)}(t_1,t_2)|^2,
\end{equation}
valid for thermal states, where $\tilde{g}_{ss}^{(2)}$ is the integrand in Eq.(\ref{eq:g2_definition}) and $g_{ss}^{(1)}$ is the normalized first-order correlation, defined as 
\begin{equation}
    g_{ss}^{(1)}(t_1,t_2) = \frac{\langle [a_s^{out}(t_1)]^{\dagger} a_s^{out}(t_2)  \rangle}{\sqrt{\langle [a_s^{out}(t_1)]^{\dagger} a_s^{out}(t_1) \rangle \langle [a_s^{out}(t_2)]^{\dagger} a_s^{out}(t_2) \rangle}}, \label{eq:g1_normalized}
\end{equation}
one can write
\begin{equation}
\begin{aligned}
        | \langle [a_s^{out}(t_1)]^{\dagger} a_s^{out}(t_2)\rangle|^2  &=\langle [a_s^{out}(t_1)]^{\dagger} [a_s^{out}(t_2)]^{\dagger} \\ &a_s^{out}(t_2) a_s^{out}(t_1) \rangle  - \langle n_s(t_1) \rangle \langle n_s(t_2) \rangle
\end{aligned}
\end{equation}
and introduce the phase-insensitive  correlation function $\tilde{G}_{ss}(\tau)= \int dt |\langle [a_s^{out}(t)]^{\dagger} a_s^{out}(t +\tau ) \rangle|$. Clearly the definitions of $\tilde{G}_{ss}^{(1)}$ and $G_{ss}^{(1)}$ do not coincide because of the loss of phase information in $\langle [a_s^{out}(t)]^{\dagger} a_s^{out}(t +\tau )\rangle$, but $\tilde{G}_{ss}^{(1)}$ can be readily measured with the setup in Fig.\ref{Fig_0}(b), since it depends only on coincidence and single photon probabilities.
%By doing the substitution, we find
%\begin{equation}
%    \frac{\langle a^{\dagger}(t_1) a^{\dagger}(t_2) a(t_2) a (t_1) \rangle}{\langle a^{\dagger}(t_1) a(t_1) \rangle \langle a^{\dagger}(t_2) a(t_2) \rangle} = 1 + \left\lvert \frac{\langle a^{\dagger}(t_1) a(t_2)\rangle}{\sqrt{\langle a^{\dagger}(t_1) a(t_1) \rangle \langle a^{\dagger}(t_2) a(t_2) \rangle}}\right\rvert^2
%\end{equation}
%By simplifying the common denominator and using the definition of $\langle n(t) \rangle$, we obtain
%This is an interesting form because we can measure the first term as it is proportional to the probability of having a coincidence between the two signals
%\begin{equation}
%    C_{ss} = \langle a^{\dagger}(t_1) a^{\dagger}(t_2) a(t_2) a (t_1) \rangle
%\end{equation}
%Additionally, the average photon number at times \(t_1\) and \(t_2\), given by \(\langle n(t_1) \rangle\) and \(\langle n(t_2) \rangle\), are experimentally accessible. By subtracting the product \(\langle n(t_1) \rangle \langle n(t_2) \rangle\) from \(C_{ss}\), we can isolate the square modulus of the first-order correlation function:
%\begin{equation}
%    |\langle a^{\dagger}(t_1) a(t_2) \rangle|^2
%\end{equation}
%Introducing the change of variables \(t_1 \rightarrow t\) and \(t_2 \rightarrow t + \tau\), and integrating over \(t\), yields a quantity which does not depend on $t$.
%\begin{equation}
%    |G_{ss}^{(1)} (\tau)|^2 = \int dt %|\langle a^{\dagger}(t) a(t +\tau ) \rangle %|^2
%\end{equation}
%The Fourier transform of $G_{ss}^{(1)} (\tau)$ provides the spectral line shape of the emitted photons \cite{Strongly_Zachary}.
We measured $\tilde{G}_{ss}^{(1)} (\tau)$ for different input powers and different values of $\Delta_p$, setting the pulse duration to $T = 5000$ ps in all the measurements.
In Fig.\ref{Fig_5}(a), we report the experimental data for $\Delta_p = 0$ and for three distinct input power levels. At low pump power (40 pJ), \(\tilde{G}_{ss}^{(1)}(\tau)\) exhibits a single, smooth peak. As the input power increases, the peak becomes progressively narrower. At the highest pump power (1200 pJ), a broad pedestal emerges around the sharp central peak. \\
Fig.\ref{Fig_5}(b) shows instead the experimental results for $\Delta_p = \Delta_{opt}$. In this configuration, the temporal correlation function maintains a single-peaked structure for increasing powers (pulse energies greater than $320$ pJ could not be investigated due to the crossing of parametric oscillation threshold). 
%Numerical simulations were conducted using the \texttt{QuTiP} function \texttt{correlation\_2op\_2t}, which directly computes \(\langle a^{\dagger}(t) a(t + \tau) \rangle\). From this, the experimentally accessible quantity \(|\langle a^{\dagger}(t) a(t + \tau) \rangle|^2\) is obtained by marginalizing over \(t\) and taking the modulus squared. The Fourier transform of \(\langle a^{\dagger}(t) a(t + \tau) \rangle\) allows for reconstruction of the emission spectrum.\\
We support the experimental observations with numerical simulations, which are shown in Fig.\ref{Fig_5}(c) for $\Delta_p = 0$ and in Fig.\ref{Fig_5}(d) for $\Delta_p = \Delta_{opt}$. In both cases, the evolution of the shape of $\tilde{G}_{ss}^{(2)}$ at increasing input powers qualitatively follows the same trend shown in Fig.\ref{Fig_5}(a,b), but the simulations systematically predict broader correlation functions. We believe that the narrower temporal correlation observed in the experiment could be related to spontaneous Raman scattering, whose effect is not included in the simulation. Indeed, noise photons produced by Raman scattering are expected to be spectrally broader than those generated by SFWM, which happens because the efficiency of the process is not limited by energy-conservation. Therefore, when Raman noise superimposes to the squeezed light generated by SFWM, the coherence time of the signal/idler mode is reduced. \\
%A complete observation of the Fourier transform of the spectrum would require measurement of \(G_{ss}^{(1)}(\tau)\) (with phase information) rather than just its modulus squared. Nonetheless, this qualitative behavior of \(|g_{ss}^{(1)}(\tau)|^2\) aligns with the simulation predictions in \ref{Fig_5}(b,d).
To elucidate the origin of the different behavior of $\tilde{G}_{ss}^{(1)}$ for $\Delta_p=0$ and $\Delta_p=\Delta_{opt}$, we recall the theoretical results derived in \cite{Strongly_Zachary,kim2025simulating} for both CW and pulsed regime. In these works, the authors showed that when the resonator is pumped at its cold resonance wavelength ($\Delta_p = 0$) and the pump power is increased, the single photon spectrum exhibits a transition from a single lorentzian peak to a doublet structure \cite{Strongly_Zachary} as a consequence of SPM and XPM induced resonance shifts. Conversely, when operating at $\Delta_p = \Delta_{opt}$, the spectral profile is predicted to rigidly shift, but retaining its Lorentzian shape even in the strongly driven regime. The very same trend in retrieved our simulations, which are shown in Fig.\ref{Fig_5}(e) for $\Delta_p = 0$ and in Fig.\ref{Fig_5}(f) for $\Delta_p = \Delta_{opt}$. 
%It is possible to see that by increasing the input power the spectrum splits for $\Delta_p = 0$ but does not when pumped at the optimal detuning.
According to the Fourier transform relation between the single photon spectra and $G_{ss}^{(1)}$, the frequency-splitting manifests as an oscillatory (ripple) features in the $G_{ss}^{(1)}$ around the main correlation peak at $\tau=0$ \cite{guidry2022quantum}. Through numerical simulations, we found that the $\tilde{G}_{ss}^{(1)}$ function follows the envelope of these oscillating background, which appears in Fig.\ref{Fig_5}(a,c) as a broad pedestal around the main peak at $\tau=0$.
Therefore, we were able to indirectly observe the strongly-driven spectral splitting from temporal signatures from the easily accessible correlation function $\tilde{G}_{ss}^{(1)}$.

%The simulations prove that the splitting in the spectrum is also expected in the pulsed regime for pulses with a sufficient length. In this case, we show that $5000$ ps are enough to observe this behavior. The experiments being in good agreement with the simulations prove that the \(|G_{ss}^{(1)}(\tau)|^2\) is modified in the high gain regime for $\Delta_p = 0$. 

\subsection{Joint temporal intensity}
The last quantity we characterized are the joint correlations between the signal and the idler mode. The correlations between the frequency of the signal and the idler photon are captured by the joint spectral amplitude (JSA) $J(\omega_s,\omega_i)$, whose modulus square gives the probability to generate a signal-idler pair at frequency $\omega_{s}$ and $\omega_{i}$ respectively \cite{caspani2017integrated}. As before, we discretize the signal and idler spectral intervals in $N$ steps of size $\Delta\omega$, such that $\omega_{s,j}=\lceil \omega_{s}\rceil+j\Delta\omega$ and $\omega_{i,k}=\lceil \omega_{i}\rceil+k\Delta\omega$ (with $j$ and $k$ ranging from $0$ to $N-1$). The JSA is thus sampled on a $N\times N$ grid to form the matrix $J_{jk}=J(\omega_{s,j},\omega_{i,k})$. 
In the low-gain regime, where the emission of multiple pairs is negligible, the state $\ket{\Psi_{LG}}$ at the point marked by a star in Fig.\ref{Fig_0}(a) can be written as \cite{BEYOND}
\begin{equation}
\ket{\Psi_{LG}} = \sqrt{(1-\beta)}\ket{\textrm{vac}}+\beta\sum_{j,k=0}^{N-1}\bar{J}_{jk}\ket{\omega_{s,j}\omega_{i,k}}, \label{eq:stato_lg}   
\end{equation}
where $|\beta|^2$ is the pair generation probability, $\bar{\mathbf{J}}$ is the low-gain JSA and $\ket{\omega_{s,j}\omega_{i,k}}=b^{\dagger}_{s,j}b^{\dagger}_{i,k}\ket{\textrm{vac}}$. Thus, the complex matrix $\mathbf{\bar{J}}$ fully describes the quantum state. A frequency-resolved coincidence measurement directly yields $|\bar{J}_{jk}|^2$, since the number of joint detections $N_{si}$ can be expressed using Eq.(\ref{eq:stato_lg}) as input state as $N_{si}=\langle a^{out\,\dagger}_{s,j}a^{out\,\dagger}_{i,k}a^{out}_{s,j}a^{out}_{i,k}\rangle\sim\eta_s\eta_i\langle b^{\dagger}_{s,j}b^{\dagger}_{i,k}b_{s,j}b_{i,k}\rangle\sim\eta_s\eta_i|\bar{J}_{jk}|^2$ (see Fig.\ref{Fig_0})(a)), where $\eta_{s(i)}$ are the signal(idler) loss from the point denoted with a star in Fig.\ref{Fig_0}(a) to the detectors (including the detection efficiency).\\
In the high-gain regime one can not neglect the contributions of multiple pairs. For example,  the state at the point marked by a star in Fig.\ref{Fig_0}(a) is a multimode-squeezed state described by the tensor product of broadband frequency modes - the Schmidt modes of the system - characterized by the squeezing parameters $\mathbf{\xi}$. One can write this state as \cite{christ2013theory,BEYOND}
\begin{equation}
    \ket{\Psi_{HG}} = \exp \left ( \sum_{j=0}^{N-1}\sum_{k=0}^{N-1}J_{jk}b_{s,j}^{\dagger}b_{i,k}^{\dagger} - \textrm{h.c}\right )\ket{\textup{vac}}, \label{eq:JSA}
\end{equation}
\begin{figure}[]
    \includegraphics[width=\linewidth]{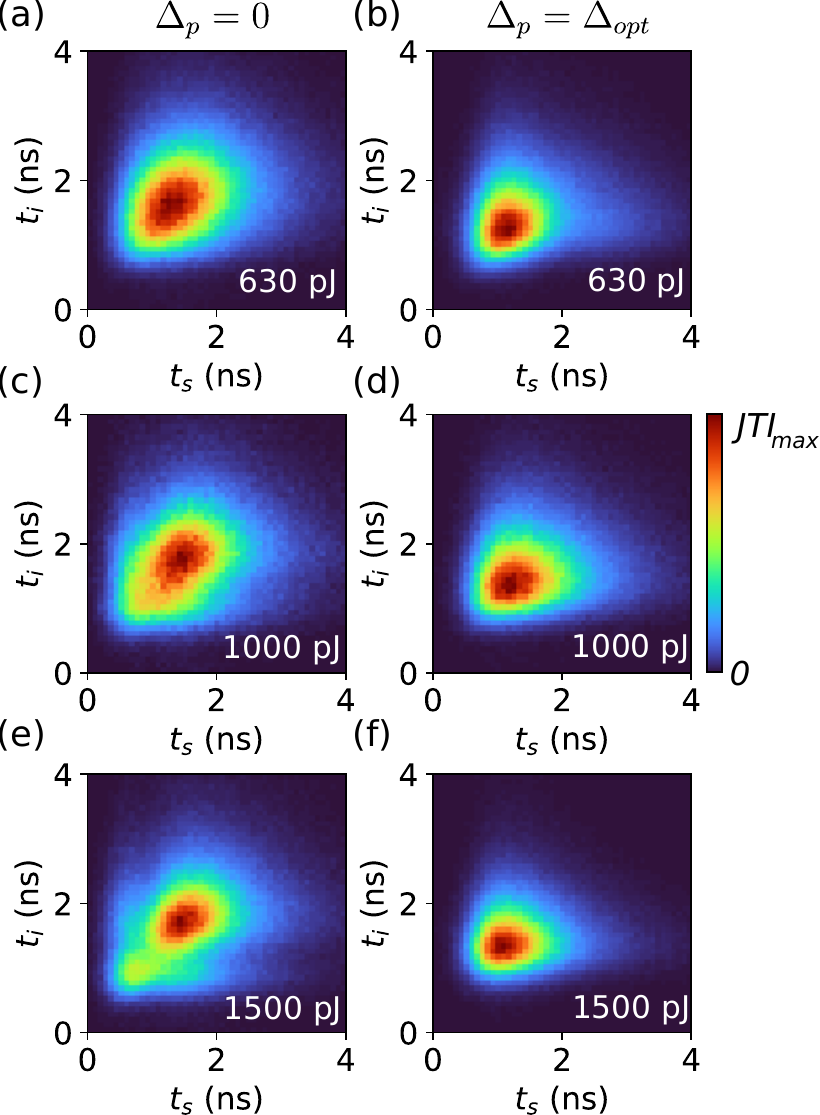}
    \caption{Two-dimensional histograms showing the time resolved coincidence measurements between the signal and the idler photons, which in the limit of infinitely low gain correspond to the JTI. Panels (a), (c) and (d) have $\Delta_p=0$, while panels (b), (d) and (f) have $\Delta_p=\Delta_{opt}$. In both configurations, the pump pulse energies are $630$ pJ, $1000$ pJ and $1500$ pJ.
    }
    \label{Fig_6}
\end{figure}
\begin{figure*}[]
    \includegraphics[scale = 0.75]{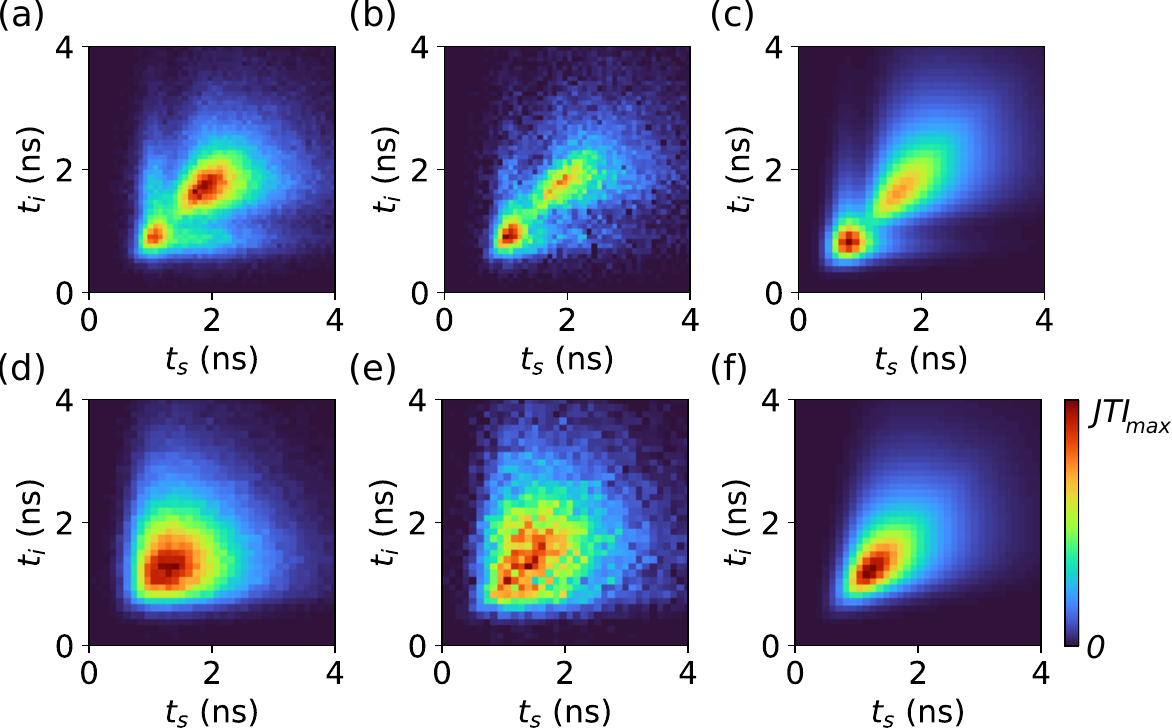}
    \caption{(a) Two-dimensional histogram showing the time resolved coincidence measurement between the signal and the idler photons. The pump pulse energy is set to $1650$ pJ and $\Delta_p=0$. (b) Approximate JTI obtained from panel (a) by applying the correction in Eq.(\ref{eq:pqp_optimal_alpha}), that uses four-photon events ($\alpha_{opt}=-35$ and $\langle n_s\rangle=1.8$). (c) Simulated JTI at a pump pulse energy of $1650$ pJ and $\Delta_p=0$. Panels (d-e-f) follows the same order and have the same description, but are related to a pump energy of $800$ pJ and $\Delta_p=\Delta_{opt}$ ($\langle n_s\rangle = 3$, $\alpha_{opt}=-8$) 
    }
    \label{Fig_7}
\end{figure*}
where $\mathbf{J}=\frac{1}{2}\mathbf{P}_s\bm{\xi}\mathbf{P}_i^*$ and $\mathbf{P}_{s(i)}$ are the matrices defining the upper-left ($\mathbf{P}_s$) and lower-right ($\mathbf{P}_i^*$) $N\times N$ diagonal blocks of the orthogonal matrix $\mathbf{P}$ in Fig.\ref{Fig_0}(a) (see Eq.(\ref{eq:JSA_continuos}) of Appendix A and Ref.\cite{kim2025simulating}).  Eq.(\ref{eq:stato_lg}) represents the first-order term of the Taylor expansion of Eq.(\ref{eq:JSA}), which allows one to term $\mathbf{J}$ as the arbitrary-gain joint spectral amplitude. Therefore, as long as pump depletion terms are neglected in Eq.(\ref{H_non_lineare}), the gaussian state in Eq.(\ref{eq:JSA}) is completely determined by $\mathbf{J}$.     
However, it is far more challenging to measure $\mathbf{J}$ for high-gains because $N_{si}$ is no more proportional to $|J_{jk}|^2$ due to multiple pair emission. The same argument holds for the relation between the number of coincidence detections $\tilde{N}_{qp}$ at times $t_{s,q}$ and $t_{i,p}$ and the joint temporal intensity $|\mathbf{\tilde{J}}|^2$ (JTI), being $\mathbf{J}$ and $\mathbf{\tilde{J}}$ related by a Fourier transform.   Strategies for measuring the arbitrary gain JSA based on stimulated emission have been theoretically reviewed in \cite{BEYOND} and experimentally validated for broad-band parametric-down conversion in periodically poled crystals in \cite{triginer2020understanding}. In principle, this approach could be extended to SFWM in a resonator, but the absence of ultra-narrowband spectral filters prevents us from implementing this method. Consequently, we focused on measuring $\mathbf{\tilde{J}}$ by time-resolved photodetection.
In Fig.\ref{Fig_6} we report $\tilde{N}_{qp}$ for increasing pump pulse energies and for both $\Delta_p=0$ and $\Delta_p=\Delta_{opt}$. In the first case, we observe a transition from a distribution with a single peak at low energy ($630$ pJ - Fig.\ref{Fig_6}(a)) to a bimodal distribution at high energy ($1500$ pJ - Fig.\ref{Fig_6}(e)). In contrast, when $\Delta_p=\Delta_{opt}$, the shape does not vary significantly with increasing power, and the distribution maintains a well-defined maximum. The behavior for $\Delta_p = 0$ and $\Delta_p = \Delta_{\text{opt}}$ is consistent with the results of $\langle n_s(t) \rangle$ shown in Fig.\ref{Fig_1}. This is expected, since in the low-gain regime $\langle n_s(t) \rangle \sim \int |\tilde{J}(t_s, t_i)|^2  dt_i$. As before, the differences between $\Delta_p=0$ and $\Delta_p = \Delta_{opt}$ can be attributed to the distinct energy mismatches $\Delta\omega$ associated with each detuning condition. A clear indication that that $\tilde{N}_{qp}$ is no longer a reliable measure of $\tilde{J}_{qp}$ comes from the estimated upper bound of $g^{(2)}_{ss}$. When this bound is calculated using the approximation $\tilde{J}\sim\sqrt{\tilde{N}_{si}}$ \cite{vernon2017truly} for $\Delta_p = 0$, the resulting values are $1.98(1)$ at 630 pJ, $1.97(1)$ at 1000 pJ, and $1.95(1)$ at 1500 pJ. These estimates significantly exceed the values obtained from direct measurements of the integrated second-order correlation of the marginal signal beam at comparable powers, as shown in Fig.\ref{Fig_4}(c). The underestimation of the Schmidt number is due to the contamination of multiple pairs to the time-resolved coincidences $\tilde{N}_{qp}$. Indeed, one can write the probability $p_{pq}\propto \tilde{N}_{qp}$ of detecting a pair of signal-idler photons at times $t_{s,q}$ and $t_{i,p}$ as
\begin{equation}
    p_{qp} = \sum_{n=1}^{\infty}h_{n}^{(1)}(\eta_s,\eta_i)P_n(t_{s,q},t_{i,p}), \label{eq:time_resolved_prob}
\end{equation}
where $h_{n}^{(1)}(\eta_s,\eta_i)$ is the probability that, given the generation of $n$ photon pairs, at least one signal and one idler photon are detected after passing through separate channels with transmittivity $\eta_s$ and $\eta_i$ (see Appendix A for the explicit expression). The quantity $P_n(t_{s,q},t_{i,p})$ is the marginal probability to generate a signal photon at time $t_{s,q}$ and an idler photon at time $t_{i,p}$, conditioned on the generation of $n$ pairs. This quantity can be calculated by expanding the exponential in Eq.(\ref{eq:JSA}), which gives
\begin{equation}
P_n(t_{s,q},t_{i,p})=\sum_{\bm{k}_n}|\textrm{Perm}(\tilde{J}_{(\bm{k}_{n,q,p})})|^2, \label{eq:perm}
\end{equation}
where $\bm{k}_n$ is a vector of length $2(n - 1)$ containing the indices corresponding to the emission times of the remaining $n - 1$ signal and $n - 1$ idler photons, while $q$ and $p$ are the indices associated with the emission times $t_{s,q}$ and $t_{i,p}$, respectively. The notation $\tilde{J}_{(\bm{k}_n,q,p)}$ refers to the submatrix of $\tilde{J}$ formed by selecting the $n$ rows and $n$ columns corresponding to the signal and idler indices specified in the vector $(\bm{k}_n, q, p)$ - that is, the $n-1$ signal and idler indices from $\bm{k}_n$, together with the additional indices $q$ and $p$. The quantity $\textrm{Perm}(\tilde{J}_{(\bm{k}_n,q,p)})$ is the $2n$-photon wavefunction, whose modulus squared describes the probability of generating $n$ pairs of signal-idler photons at the $2n$ times specified by the indices in $(\bm{k}_n,q,p)$. It follows that $P_1(t_{s,q},t_{i,p})=|\tilde{J}|_{qp}^2$. Being a bosonic field, $P_n$ is invariant under any permutation of signal(idler) photons - corresponding to rows(column) swaps- which is reflected in the symmetry properties of the permanent function. The sum in Eq.(\ref{eq:perm}) runs over all the indices except $(q,p)$. This marginalization process over an increasing number of photons progressively masks the correlations in the emission time of each pair of signal-idler photons described by $\tilde{J}$. Indeed, as the number of photon pairs $n$ increases, there is an increasing probability that the signal and the idler photons detected at time $t_{s,q}$ and $t_{i,p}$ belong to different pairs, hence to uncorrelated emissions. To support this intuition, we computed $P_n$ up to $n=10$ through Monte Carlo simulations, and we evaluated its fidelity with the product of the marginalized probabilities $P_n^{(s)}(t_s)=\int P_n(t_s,t_i)dt_i$ and $P_n^{(i)}(t_i)=\int P_n(t_s,t_i)dt_s$. This fidelity increases monotonically with $n$ (see Appendix A). \\
Moreover, the similarity between successive distributions also increases with $n$, as indicated by a decreasing energy distance between $P_n$ and $P_{n+1}$ for large $n$ \cite{szekely2004testing}, suggesting  convergence toward a limiting distribution (see Appendix A). This indicates a strategy for approximately recovering the time correlations described by $P_1(t_{s,q},t_{i,p})=|\tilde{J_{qp}}|^2$ from measurements of $p_{qp}$ and the time-resolved four-fold coincidences $p_{qpmn}$. We report here the final result, leaving the full derivation to Appendix A.
The time correlations $P_1$ can be approximated, up to corrections arising from triple and higher order pair contributions, as
\begin{equation}
    P_1(t_{s,q},t_{i,p}) = \left (p_{qp}-\alpha_{opt}\sum_{mn}p_{qpmn} \right)[h_{1}^{(1)}]^{-1}+\mathcal{O}(\textrm{triples}),\label{eq:P_1_approx_simple}
\end{equation}
where $\alpha_{opt}$ is a scaling constant that depends on both the signal/idler loss from generation to detection and the average photon number (see Appendix A).
We use Eq.(\ref{eq:P_1_approx_simple}) to approximate $\mathbf{\tilde{J}}$ from the measured time-resolved two-fold and four-fold coincidence probabilities $p_{qp}$ and $p_{qpmn}$, for two different configurations. We first set $\Delta_p=0$ and the pump energy to $1650$ pJ, and show $p_{qp}$ in Fig.\ref{Fig_7}(a). If this distribution is erroneously interpreted as  $|\mathbf{\tilde{J}}|^2$, the corresponding upper bound to the purity would be $0.92(1)$. In contrast, the upper bound calculated from the simulation, shown in Fig.\ref{Fig_7}(c), is $0.75$. We then recorded the time-resolved four-photon coincidences $p_{qpmn}$, and used them in Eq.(\ref{eq:P_1_approx_simple}) to approximate $|\tilde{J}_{qp}|^2$. %(note that the value of $\alpha_{opt}$,  calculated from Eq.(\ref{eq:alpha_opt}), uses the transmission coefficients $\eta_s$ and $\eta_i$ measured for the signal and the idler photons to compute the $h$-coefficients, and by setting $r_n = (1-\lambda^2)\lambda^{2n}$, where $\lambda\sim\tanh(\textrm{arcsinh}(\sqrt{\langle n_s\rangle }))$ ($\langle n_s\rangle \sim 1.8$ in this experimental configuration) \cite{BEYOND}. It is worth to stress that this expression holds only for a two-mode squeezed state, which may deviate from the photon number distribution of the state in Eq.(\ref{eq:JSA}) if the JTA is not factorable \cite{mauerer2009colors}. However, for low Schmidt numbers, as in our case, it constitutes very good approximation. 
The result of the four-photon correction is shown in Fig.\ref{Fig_7}(b). Compared to the raw coincidence measurement shown in Fig.\ref{Fig_7}(a), the JTI is now qualitatively more similar to the simulation in Fig.\ref{Fig_7}(c), but most importantly reveals a time correlation between the signal and the idler photons that was originally blurred in Fig.\ref{Fig_7}(a) by multi-photon contamination. This observation is quantitatively supported by the decrease of the upper bound to the purity, which is now $0.79(2)$. In a second configuration, we set the pump power to $800$ pJ and $\Delta_p=\Delta_{opt}$, and show the raw two-fold coincidences in Fig.\ref{Fig_7}(d), the JTI - approximated by using four-photon corrections - in Fig.\ref{Fig_7}(e), and the simulation in Fig.\ref{Fig_7}(f). The average number of photons in this configuration is $\langle n_s\rangle\sim3$. The upper bound to the purity is $0.99(1)$ in Fig.\ref{Fig_7}(d), $0.94(1)$ in Fig.\ref{Fig_7}(e) and $0.89$ in Fig.\ref{Fig_7}(f). Also in this case, the four-photon correction brought up correlations which were hard to detect in Fig.\ref{Fig_7}(d), but clear in the simulation. The discrepancy between the theory and the experiment is due to triple and higher order pair contributions, which impact since $\langle n_s\rangle$ is high. \\
Discrepancies could be further reduced by incorporating in Eq.(\ref{eq:P_1_approx_simple}) a correction term which is proportional to the time-resolved six-fold detections (see Appendix A), yet this would require measuring the six-fold detections with a high signal-to-noise ratio to avoid error-amplification. The low rates of six-fold events prevented us to collect sufficient statistics, hence we limited our analysis only to four-photon correction.
\section{Discussion}
\begin{figure}[t!]
\includegraphics[width=\linewidth]{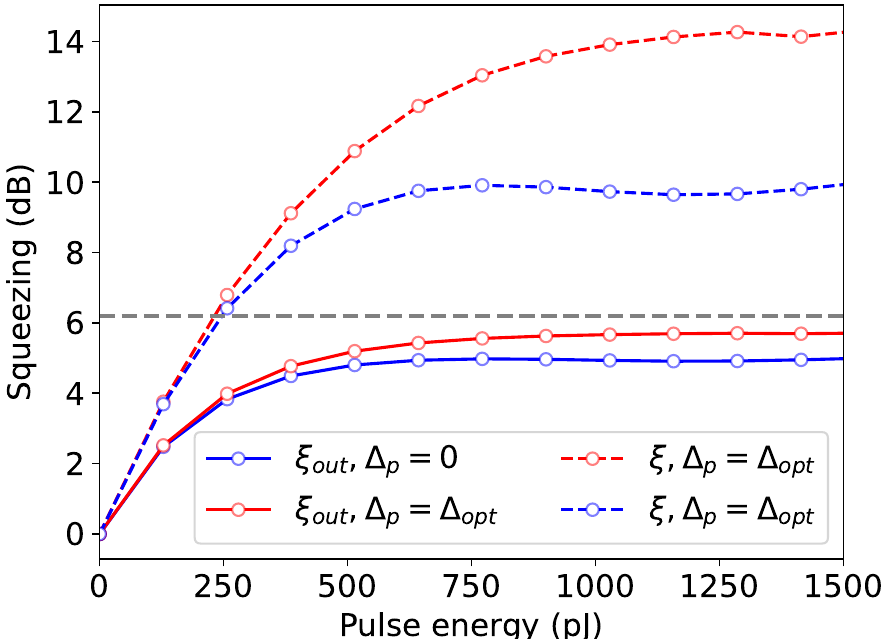}
    \caption{Quadrature squeezing of the dominant Schmidt supermode as a function of the pump pulse energy. The dashed lines correspond to the \emph{internal} squeezing $\xi$, while the solid lines to the squeezing at the output waveguide $\xi_{out}$. The gray dashed line is the upper bound $\xi_{out}^{max}$ set by the finite escape efficiency of the resonator.
    }
    \label{Fig_9}
\end{figure}

One of the key findings from the previous sections is that, for each value of the pump pulse energy, there exists an optimal pump frequency detuning that simultaneously maximizes both the output photon flux and the the spectral purity of the marginal beams. Specifically, within the interval shown in Fig.\ref{Fig_2}(a,b) and at $\Delta=\Delta_{opt}$, the generated photon flux grows quadratically with the pulse energy. However, this trend contrasts markedly with that of quadrature squeezing $\xi_{\lambda,out}$ observed in each of the $N$ squeezed supermodes $\lambda$ at the output waveguide. The output squeezing $\xi_{\lambda}^{out}$ is related to the \emph{internal} squeezing amplitudes $\bm{\xi}$ as
\begin{equation}
    \xi_{\lambda,out}=-\frac{1}{2}\ln\left (1-p_e+p_e e^{-2\xi_{\lambda}}\right),
\end{equation}
which implies an upper bound:
$\xi_{out}^{max}=-\frac{1}{2}\ln\left (1-p_e\right)\sim 6.2$ dB.
Using the methods described in Appendix A, we simulated $\mathbf{\xi}$ and $\mathbf{\xi}_{out}$ as functions of the pump pulse energy for both $\Delta_p=0$ and $\Delta_p=\Delta_{opt}$. The results, shown in Fig.\ref{Fig_9} for the dominant squeezed supermode indicate that $\xi_{out}$ saturates at $5$ dB for $\Delta_p=0$, and at $5.7$ dB for $\Delta_p=\Delta_{opt}$, i.e., a modest improvement with pump detuning. Notably, this performance gap is expected to widen with increasing escape efficiency \cite{kim2025simulating}). In contrast, the \emph{internal} squeezing $\xi$ reaches up to $14$ dB for $\Delta_p=\Delta_{opt}$, over $4$ dB higher than what is achievable at $\Delta_p=0$. This suggests that combining a detuned pump with higher escape efficiency could significantly enhance the extraction of quadrature squeezing from the resonator. In addition, optimizing the pump duration or temporal shape -such as introducing a frequency-chirp- could further help saturate the bound $\xi_{out}^{max}$ set by the escape efficiency.\\ 
Finally, it is worth noting that in the case of dual-pump SFWM \cite{zhang2021squeezed}, the optimal strategies may differ from those discussed here. This is due to the distinct effects of SPM and XPM on the two pumps and on the degenerate squeezed modes. Exploring how key source metrics, such as squeezing strength, Schmidt number and state purity, depend on the detuning, duration and shape of both pumps would be highly valuable for CV quantum computing applications \cite{aghaee2025scaling,larsen2025integrated}.

\section{Conclusion}
In this work, we presented a comprehensive characterization of pulsed squeezed light generation in a silicon nitride microresonator, focusing on the transition from the low to the high-parametric gain regime, where SPM, XPM, and time-ordering corrections in the squeezing Hamiltonian cannot be neglected.

We observed several characteristic indicators of the high-gain regime, including a local maximum in the signal/idler generation rate as the input pump power varied, high spectral purity under detuned pump excitation, and the appearance of a broad pedestal in the first-order coherence function, indicative of spectral splitting in the single-photon spectra caused by self-phase and cross-phase modulation (SPM/XPM). All the experimental results are supported by theoretical simulations, aligning very well with the observations.\\
Our main finding is that, for a resonator with anomalous dispersion, for each pump power there is an optimal pump frequency detuning that maximizes the generation rate and simultaneously leads to high spectral purity. We also found that, at the optimal detuning, pulse durations longer than the cavity lifetime increase the generation rate for a fixed pulse energy without significantly reducing the spectral purity.
As a final step, we showed that time-resolved coincidence measurements are ineffective for estimating the JTI in the high-gain regime due to multi-pair contamination. We addressed this problem by proposing and validating an error-correction strategy that uses the marginal probability distributions of time-resolved multi-photon events to correct coincidence measurements, allowing us to recover, to some approximation, the time correlations predicted by the JTI.
Our results deepen the understanding of microresonators as sources of non-classical light in a regime of great relevance for continuous-variable quantum computing and sensing, providing a practical strategy for optimizing and characterizing their performance.

\section*{Acknowledgment}
D.B. acknowledges the support of Italian MUR and the European Union - Next Generation EU through the PRIN project number F53D23000550006 - SIGNED. E.B., M.B., M.G. and M.L. acknowledge the PNRR MUR project PE0000023-NQSTI. All the authors acknowledge the support of Xanadu Quantum Technology for providing the samples.
%\subsection*{Author contributions}
%M.B conceived the original idea. E.B. and M.B performed the experiment, analyzed the data and wrote the manuscript. All the authors reviewed the manuscript and provided useful discussions. M.B, D B., M.G and M.L. supervised the whole work.
%\subsection*{Competing interests}
%The authors declare no competing interests.
%\subsection*{Data availability}
%Data is available from the corresponding author upon reasonable request.
\section*{Appendix A: High-order pair contribution to coincidence measurements}
\subsection*{Simulation of the arbitrary gain JTA through the master equation}
The simulation of the JTA is performed using the open-access Python library \texttt{QuTiP}. To accomplish this, we first calculate the second-order moment \mbox{$M(t,t')=\langle b_s(t)b_i(t')\rangle=2\frac{\gamma_e}{p_e}\langle c_s(t)c_i(t')\rangle$} \cite{vernon2019scalable} by solving Eq.(\ref{eq:master_equation}) for $\rho(t)$ and by evaluating $\langle c_s(t)c_i(t')\rangle$ using the \texttt{QuTiP.correlation\_2op\_2t} function \cite{QuTiP}. In practice, $M(t,t')$ is sampled over a finite $N\times N$ grid of points separated by the discretization step $\Delta t$, i.e. we calculate the matrix elements $M_{pq}=M(t_0+p\Delta t,t_0+q\Delta t)$, where $t_0$ is an arbitrary offset. We then use the Williamson's theorem and the Bloch-Messiah decomposition to formally write a joint decomposition of the second-order moment as \cite{vernon2019scalable}
\begin{equation}
    M(t,t')=\sum_{\lambda} \frac{\sinh{(\xi_\lambda)}}{2}f^{(s)}_{\lambda}(t)f^{(i)}_{\lambda}(t'),
\end{equation}
and connect the functions $f^{(s,i)}_{\lambda}(t)$ and the squeezing parameters $\xi_{\lambda}$ to the singular value decomposition $M_{qp}=\sum_{\lambda}D_{\lambda}P^{(s)}_{\lambda q}(P^{(i)}_{\lambda p})^*$ using the relations \cite{mejia2015very}
\begin{equation}
\label{eq:connections}
\begin{aligned}
D_{\lambda} &= \frac{\sinh(2\xi_{\lambda})}{2\Delta t}, \\
P^{(s)}_{\lambda q} &= \sqrt{\Delta t}f^{(s)}_{\lambda}(t_0 + q\Delta t), \\
P^{(i)}_{\lambda p} &= \sqrt{\Delta t}(f^{(i)}_{\lambda})^*(t_0 + p\Delta t).
\end{aligned}
\end{equation}
\\
\noindent The columns of the matrices $\mathbf{\tilde{P}}^{(s)}$ and $(\mathbf{\tilde{P}}^{(i)})^*$ are the Fourier transform of the respective columns  of the matrices $\mathbf{P}^{(s)}$ and $\mathbf{P}^{(i)}$, i.e, they represent the squeezed temporal modes. The multimode squeezed state at the point denoted by a star in Fig.\ref{Fig_0}(a) is given by \cite{vernon2019scalable}
\begin{equation}
    \ket{\Psi_{HG}}=\exp \left( \int\tilde{J}(t,t')b^{\dagger}_s(t)b^{\dagger}_i(t') - \textrm{h.c} \right)\ket{\textrm{vac}}, \label{eq:JSA_continuos}
\end{equation}
where  $\tilde{J}(t,t')=\sum_{\lambda}\frac{\xi_{\lambda}}{2}f^{(s)}_{\lambda}(t)f^{(i)}_{\lambda}(t')$ and we used the same notation of Eq.(\ref{eq:JSA}) as the latter represents the discretized version of Eq.(\ref{eq:JSA_continuos}) expressed in the time domain. It follows that $\tilde{J}_{qp}=\sum_{\lambda}R_{\lambda}P^{(s)}_{\lambda q}(P^{(i)}_{\lambda p})^*$, where 
\begin{equation}
    R_{\lambda}= \frac{\textrm{arcsinh}(2D_{\lambda}\Delta t)}{2\Delta t}
\end{equation}
and $\xi_{\lambda}=2\Delta tR_{\lambda}=\textrm{arcsinh}(2D_{\lambda}\Delta t)$. The JTA is calculated over a square grid of $50\times50$ points, using a discretization step of $80$ ps.

\subsection*{Calculation of the marginalized time of arrival probabilities}
Once that the JTA matrix $\tilde{\mathbf{J}}$ has been calculated, Eq.(\ref{eq:perm}) can be used to evaluate the marginal probability $P_n(t_{s,q},t_{i,p})$ to generate a signal photon at time $t_{s,q}$ and an idler photon at time $t_{i,p}$ when $n$ pairs are produced. Since the marginalization involves integration over $n-2$ dimensions, the computational cost scales as $\mathcal{O}(N^{n})$ and becomes rapidly intractable to be run in a standard laptop already at $n=4$. Therefore, the sum in Eq.(\ref{eq:perm}) is computed by Monte Carlo integration. Specifically, we used the standard Metropolis-Hastings algorithm to sample from the $n$-dimensional distribution $|\textrm{Perm}(\tilde{J}_{(\bm{k}_n,q,p)})|^2$. The simulations reported in Fig.\ref{Fig_7} used Markov-chains with 80000 samples, a burn-in period of 1000 samples, and thinning-factor of 200. These parameters  were found to be sufficient to  reach a the stationary distribution and eliminate sample correlations. After integration, the samples are re-binned into a $50\times50$ grid to increase the number of samples per each bin.\\
To model the detection probabilities $h_n^{(1)}(\eta_s,\eta_i)$, we consider that each of the $n$ signal(idler) photons in the bus waveguide has a probability $\eta_{s(i)}$ of reaching a specific photodetector, and $1-\eta_{s(i)}$ of being lost due to channel losses or routed into a different output. For example, in Fig.\ref{Fig_0}(b), the number of output channels for the signal(idler) photon is $N_C=2$. Thus, the probability that exactly $k$ signal(idler) photons arrive at specific photodetector is given by the binomial distribution $\binom{n}{k}\eta_{s(i)}^{k}(1-\eta_{s(i)})^{n-k}$. From this, the probability that at least one signal and one idler photon are detected by the SNSPD threshold detectors (triggering a coincidence event) is 

\begin{align}
h_n^{(1)}(\eta_s,\eta_i) &= \sum_{k_s=1,k_i=1}^{n} \binom{n}{k_s} \binom{n}{k_i} 
\eta_s^{k_s} \eta_i^{k_i} \notag \\
&\quad \times (1-\eta_s)^{n-k_s} (1-\eta_i)^{n-k_i}.
\label{eq:h_n_combinations}
\end{align}
Similarly, this argument can be extended to express the probability $h_n^{(m)}$ that, given $n$ photon pairs are generated, at least $m$ signal and $m$ idler photon are detected, as 
\begin{equation}
\begin{split}
h_n^{(m)}(\bm{\eta}_{s},\bm{\eta}_{i}) 
&= \sum_{\mathbf{k}_s=\mathbf{1}_m,\mathbf{k}_i=\mathbf{1}_m} \Pi(\mathbf{k}_s|n,\bm{\eta}_s)\Pi(\mathbf{k}_i|n,\bm{\eta}_i) \\
&\quad \times \theta \left( \sum_{j=1}^m k_{s,j}-n \right)\theta \left( \sum_{j=1}^m k_{i,j}-n \right),
\end{split}
\label{eq:h_n_m}
\end{equation}
where $\Pi(\mathbf{k}|n,\bm{\eta})$ is the multinomial probability of getting the outcome $\mathbf{k}=(k_1,...,k_m)$ with success probabilities $\bm{\eta}=(\eta_1,...,\eta_m)$ in $n$  independent trials and $\theta$ is the Heaviside function. 
\subsection*{Energy distance test and correlations in the marginal distributions $P_n$}
\begin{figure*}[t!]
    \includegraphics[scale = 0.42]{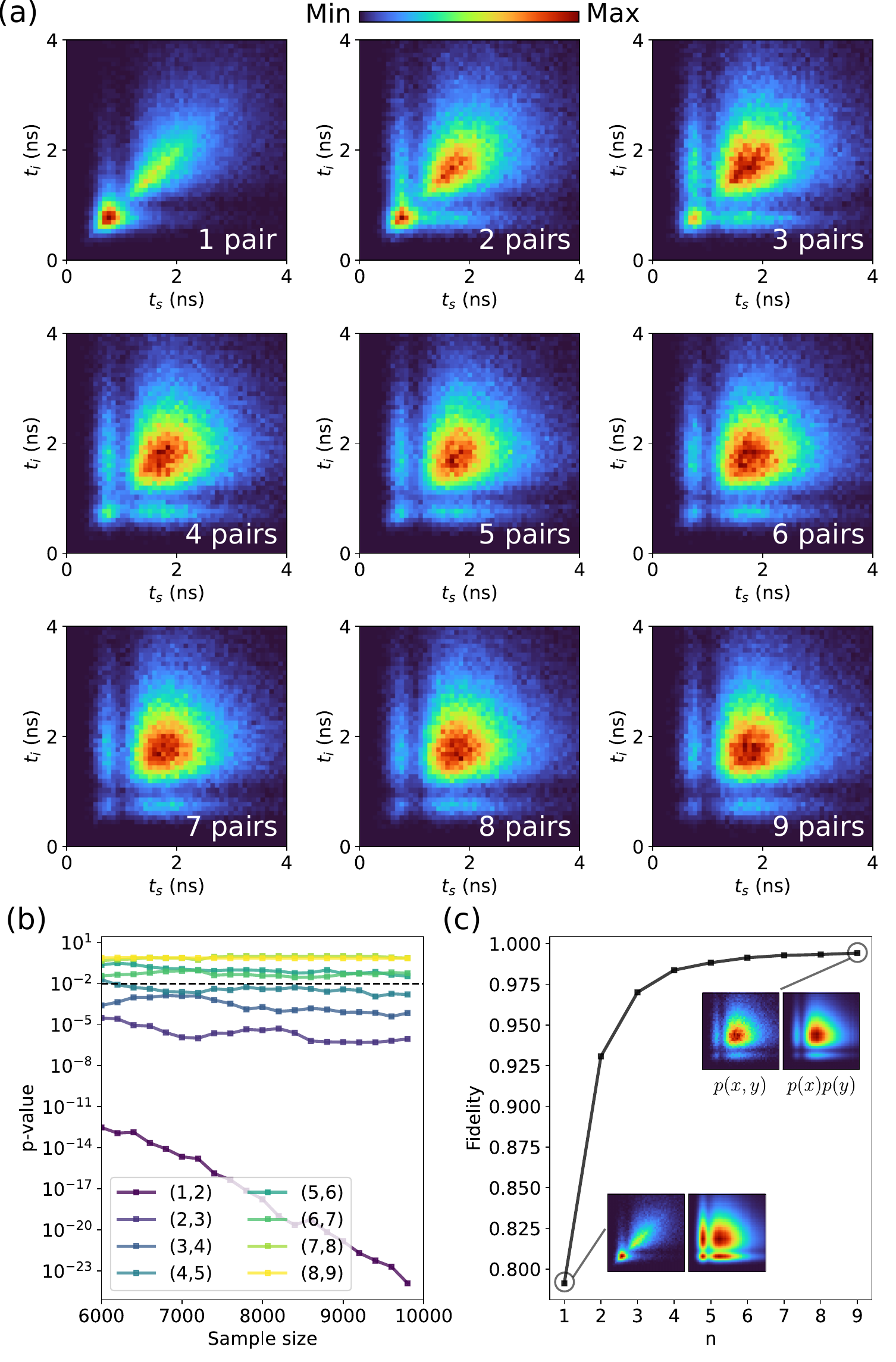}
    \caption{(a) Markov-chain Monte Carlo simulations of the probability distributions $P_n(t_s,t_i)$ in Eq.(\ref{eq:time_resolved_prob}), where $n$ labels the number of generated pairs. (b) P-values of the energy distance test of similarity between the probability distributions $(P_n,P_{n+1})$ as a function of the sample size. The dashed line marks $1\%$ of the confidence value. Above this threshold, the test is considered to fail and the two distributions are considered to be indistinguishablee. (c) Fidelity between $P_n(t_s,t_i)$ and $P^{(s)}_n(t_s)P^{(i)}_n(t_i)$, where $P^{(s(i))}_n(t_{s(i)})$ is the probability distribution calculated by marginalizing $P_n(t_s,t_i)$ over the signal(idler) photon arrival time. The two insets show a comparison between these quantities for $n=1$ and $n=9$.
    }
    \label{Fig_8}
\end{figure*}
The energy distance $D^2(F,G)$ represents a statistical distance between two (generally multivariate) probability distributions $F(\mathbf{u})$ and $G(\mathbf{v})$ \cite{szekely2004testing}. For $m$ samples drawn from $F$ and $n$ from $G$, $D^2(F,G)$ can be calculated as
\begin{equation}
    D^2(F,G) = 2A-B-C, \label{eq:energy_distance}
\end{equation}
where the quantities $A$, $B$ and $C$ are given by
\begin{equation}
\begin{aligned}
A &= \frac{1}{m n} \sum_{i=1}^{m} \sum_{j=1}^{n} \| X_i - Y_j \|, \\
B &= \frac{1}{m^2} \sum_{i=1}^{m} \sum_{j=1}^{m} \| X_i - X_j \|, \\
C &= \frac{1}{n^2} \sum_{i=1}^{n} \sum_{j=1}^{n} \| Y_i - Y_j \|, \label{eq:energy_distance_definitions}
\end{aligned}
\end{equation}
and $\| \cdot\|$ is a metric distance (for example the Euclidean distance) between two samples. It can be shown that $D^2(F,G)\ge0$, where the sign of equality holds if and only if $F=G$ \cite{szekely2004testing}. From the energy distance $D^2$, one can construct a statistical permutation test to assess whether the two distributions are equal—this forms the null hypothesis, with an associated p-value. We used this test to demonstrate that the marginalized probability distributions $P_n$, defined in Eq.(\ref{eq:perm}), become progressively more similar as $n$ increases, eventually converging to a limiting distribution as $n\rightarrow \infty$.  We simulated $P_n$ for a pump detuning $\Delta_p=0$ and a pulse energy $\epsilon = 1650 $ pJ, showing the resulting distributions in Fig.\ref{Fig_7}(a) up to $n=9$. We then set $F=P_n(t_s,t_i)$, $G = P_{n+1}(t_s,t_i)$, and performed the statistical energy distance test by using the Python library \texttt{Hyppo} \cite{panda2019hyppo}. The returned values of the p-value for each pair of distributions $(P_n,P_{n+1})$ are shown in Fig.\ref{Fig_7}(b) as a function of the sample size. At a significance level of $1\%$, the null hypothesis is already accepted for $n=4$, indicating that samples from $P_4$ and $P_5$ are likely drawn from the same underlying distribution. This trend is also evident in Fig.\ref{Fig_7}(a), which shows a clear convergence toward a limiting distribution for increasing $n$, with minimal differences from $P_n$ to $P_{n+1}$ for $n\ge4$.\\
Furthermore, Fig.\ref{Fig_7}(a) reveals that the correlation between the arrival times $t_s$ and $t_i$ diminuishes as $n$ increases. We quantified the similarity between $P_n(t_s,t_i)$ and the product of the marginalized probabilities $P_n^{(s)}(t_s)=\int P_n(t_s,t_i)dt_i$ and $P_n^{(i)}(t_i)=\int P_n(t_s,t_i)dt_s$ by evaluating the fidelity $\mathcal{F}_n$ as
\begin{equation}
    \mathcal{F}_n=\frac{\int P_n(t_s,t_i)(P_n^{(s)}(t_s)P_n^{(s)}(t_s))dt_sdt_i}{\left (\int P_n(t_s,t_i)^2dt_sdt_i \int (P_n^{(s)}(t_s)P_n(t_i)^{(i)})^2 dt_sdt_i \right)^{\frac{1}{2}}}. \label{eq:fidelity}
\end{equation}
This quantity is shown in Fig.\ref{Fig_7}(c) as a function of $n$. For $n>4$, $\mathcal{F}>98\%$, while for $n=9$ the distribution $P_9$ becomes almost indistinguishable from $P_9^{(s)}P_9^{(i)}$ and $\mathcal{F}>99.9\%$.
\subsection*{JTA correction from higher-order photodetections}
The aim of this section is to derive Eq.(\ref{eq:P_1_approx_simple}) of the main text, which approximates  the time correlations $P_1(t_s,t_i)=|\tilde{J}(t_s,t_i)|^2$ up to triple pair contributions.
We start by defining $P_2(t_{s,q},t_{i,p}) = r_2\bar{p}_{qp}$, where $r_2$ is the probability of generating two pairs and $\bar{p}$ is a normalized probability distribution, and write $P_n$ as $P_n(t_{s,q},t_{i,p})= r_n(\bar{p}_{qp}+\delta \bar{p}^{(n)}_{qp})$ for $n\ge3$, where $\delta\bar{p}^{(n)}_{qp}$ is a small perturbation to $\bar{p}_{qp}$, which allows one to write Eq.(\ref{eq:time_resolved_prob}) as
\begin{equation}
    p_{qp} = h_1^{(1)}P_1(t_{s,q},t_{i,p})+\left (\sum_{n\ge2}h_n^{(1)}r_n \right )\bar{p}_{qp}+\sum_{n\ge3}r_nh_n^{(1)} \delta \bar{p}^{(n)}_{qp}, \label{eq:pqp_expanded}
\end{equation}
and the marginalized four-photon probability as
\begin{equation}
    \sum_{mn}p_{qpmn} = \left(\sum_{n\ge2}h^{(2)}_n r_n\right)\bar{p}_{qp}+\sum_{n\ge3}h^{(2)}_nr_n \delta \bar{p}^{(n)}_{qp} \label{eq:four-fold}
\end{equation}
where $h^{(2)}_n$ can be calculated from Eq.(\ref{eq:h_n_m}) and is the probability that, given the generation of $n$ photon pairs, at least two signal and two idler photon are detected. By multiplying Eq.(\ref{eq:four-fold}) by a scaling constant $\alpha$ and by adding the result to Eq.(\ref{eq:pqp_expanded}), one obtains
\begin{equation}
\begin{aligned}
    p_{qp} + \alpha\sum_{mn}p_{qpmn} 
    &= h_1^{(1)}P_1+\left (\sum_{n\ge2}r_n(h^{(1)}_n+\alpha h^{(2)}_{n}) \right)\bar{p}_{qp} \\
    &\quad + \sum_{n\ge3}r_n(h^{(1)}_n+\alpha h^{(2)}_n)\delta\bar{p}^{(n)}_{qp}. \label{eq:twofold_corrected}
\end{aligned}
\end{equation}
By setting
\begin{equation}
    \alpha=\alpha_{opt}=-\frac{\sum_n h^{(1)}_nr_n}{\sum_n h^{(2)}_n r_n}, \label{eq:alpha_opt}
\end{equation}
one can cancel the first term on the right hand side of Eq.(\ref{eq:twofold_corrected}), which then reduces to
\begin{equation}
    p_{qp}-\alpha_{opt}\sum_{mn}p_{qpmn}=h_1^{(1)}P_1+\sum_{n\ge3}r_n(h^{(1)}_n+\alpha_{opt} h^{(2)}_n)\delta\bar{p}^{(n)}_{qp}, \label{eq:pqp_optimal_alpha}
\end{equation}
which is precisely Eq.(\ref{eq:P_1_approx_simple}) of the main text, where we defined
\begin{equation}
\mathcal{O(\textrm{triples})}=-[h_1^{(1)}]^{-1}\sum_{n\ge3}r_n(h^{(1)}_n+\alpha_{opt} h^{(2)}_n)\delta\bar{p}^{(n)}_{qp}.
\end{equation}
Therefore, the error in the estimation of $P_1$ from the four-photon corrected coincidences has the leading term which depends on the contribution of triple pairs. 
We now show that by incorporating an additional correction term to Eq.(\ref{eq:twofold_corrected}), arising from six-fold coincidences, the error in the estimation of $P_1$ can be reduced from $\mathcal{O}(\delta\bar{p}_{qp}^{(3)})$ to $\mathcal{O}(\delta\bar{p}_{qp}^{(4)})$. To this end, we write the marginalized six-photon probability as
\begin{equation}
    \sum_{mnrs} p_{qpmnrs} = \left (\sum_{n\ge3} h_n^{(3)}r_n\right)\bar{p}_{qp}+\sum_{n\ge3}h_n^{(3)}r_n\delta \bar{p}_{qp}^{(n)}, \label{eq:sixfold_marginalized}
\end{equation}
where $h_3^{(n)}$ is the probability that, given $n$ photons pairs are generated, at least three signal and three idler photons are detected, and whose expression is given in Eq.(\ref{eq:h_n_m}). By multiplying Eq.(\ref{eq:sixfold_marginalized}) by a scaling constant $\beta$ and adding it to Eq.(\ref{eq:pqp_expanded}) one obtains
\begin{equation}
\begin{aligned}
    p_{qp} &+  \alpha\sum_{mn}p_{qpmn} +\beta\sum_{mnrs}p_{qpmnrs}
    = h_1P_1\\ & +\left (\sum_{n\ge2}r_n(h^{(1)}_n+\alpha h^{(2)}_{n}+\beta h_n^{(3)}) \right)\bar{p}_{qp} \\
    &+ \sum_{n\ge3}r_n(h^{(1)}_n+\alpha h^{(2)}_n+\beta h_n^{(3)})\delta\bar{p}^{(n)}_{qp}. \label{eq:sixfold_corrected}
\end{aligned}
\end{equation}
By linear inversion, one can determine values  $\alpha=\alpha_{opt}$ and $\beta=\beta_{opt}$ that cancel the second term on the right hand side, such that  $h^{(1)}_n+\alpha_{opt} h^{(2)}_n+\beta_{opt} h_n^{(3)}=0$ for $n=3$. Thus, by applying both four-fold and six-fold correction to coincidence events, the error in the estimation of $P_1$ can be improved from $\mathcal{O}(\delta\bar{p}_{qp}^{(3)})$ to $\mathcal{O}(\delta\bar{p}_{qp}^{(4)})$. This approach can be generalized to higher-order corrections: for example, by using up to $m$-fold coincidences, one can eliminate  error terms up to $\mathcal{O}(\delta\bar{p}_{qp}^{(m)})$. However, caution must be exercised in the linear inversion process. For fixed $n$, the terms $h_n^{(m)}$ become vanishingly small with increasing $m$.  Intuitively, the probability that $m$ photon pairs are detected given that $n$ pairs are generated scales approximately as $(\eta_{s}\eta_i)^m$, making the  inversion problem increasingly ill-conditioned. As a result, sampling noise in the $m$-fold coincidences used to correct $p_{qp}$ can be strongly amplified, potentially leading to artifacts in the reconstructed JTI. To mitigate this issue, methods that avoid direct inversion or incorporate additional physical constraints to the JTI, such as non-negativity and normalization of the JTI, could be employed. However, exploring such strategies is beyond the scope of this manuscript.

\bibliography{Biblio}

\begin{thebibliography}{40}%
\makeatletter
\providecommand \@ifxundefined [1]{%
 \@ifx{#1\undefined}
}%
\providecommand \@ifnum [1]{%
 \ifnum #1\expandafter \@firstoftwo
 \else \expandafter \@secondoftwo
 \fi
}%
\providecommand \@ifx [1]{%
 \ifx #1\expandafter \@firstoftwo
 \else \expandafter \@secondoftwo
 \fi
}%
\providecommand \natexlab [1]{#1}%
\providecommand \enquote  [1]{``#1''}%
\providecommand \bibnamefont  [1]{#1}%
\providecommand \bibfnamefont [1]{#1}%
\providecommand \citenamefont [1]{#1}%
\providecommand \href@noop [0]{\@secondoftwo}%
\providecommand \href [0]{\begingroup \@sanitize@url \@href}%
\providecommand \@href[1]{\@@startlink{#1}\@@href}%
\providecommand \@@href[1]{\endgroup#1\@@endlink}%
\providecommand \@sanitize@url [0]{\catcode `\\12\catcode `\$12\catcode `\&12\catcode `\#12\catcode `\^12\catcode `\_12\catcode `\%12\relax}%
\providecommand \@@startlink[1]{}%
\providecommand \@@endlink[0]{}%
\providecommand \url  [0]{\begingroup\@sanitize@url \@url }%
\providecommand \@url [1]{\endgroup\@href {#1}{\urlprefix }}%
\providecommand \urlprefix  [0]{URL }%
\providecommand \Eprint [0]{\href }%
\providecommand \doibase [0]{https://doi.org/}%
\providecommand \selectlanguage [0]{\@gobble}%
\providecommand \bibinfo  [0]{\@secondoftwo}%
\providecommand \bibfield  [0]{\@secondoftwo}%
\providecommand \translation [1]{[#1]}%
\providecommand \BibitemOpen [0]{}%
\providecommand \bibitemStop [0]{}%
\providecommand \bibitemNoStop [0]{.\EOS\space}%
\providecommand \EOS [0]{\spacefactor3000\relax}%
\providecommand \BibitemShut  [1]{\csname bibitem#1\endcsname}%
\let\auto@bib@innerbib\@empty
%</preamble>
\bibitem [{\citenamefont {Andersen}\ \emph {et~al.}(2016)\citenamefont {Andersen}, \citenamefont {Gehring}, \citenamefont {Marquardt},\ and\ \citenamefont {Leuchs}}]{squeezed_light_30_years_review}%
  \BibitemOpen
  \bibfield  {author} {\bibinfo {author} {\bibfnamefont {U.~L.}\ \bibnamefont {Andersen}}, \bibinfo {author} {\bibfnamefont {T.}~\bibnamefont {Gehring}}, \bibinfo {author} {\bibfnamefont {C.}~\bibnamefont {Marquardt}},\ and\ \bibinfo {author} {\bibfnamefont {G.}~\bibnamefont {Leuchs}},\ }\bibfield  {title} {\bibinfo {title} {30 years of squeezed light generation},\ }\href@noop {} {\bibfield  {journal} {\bibinfo  {journal} {Physica Scripta}\ }\textbf {\bibinfo {volume} {91}},\ \bibinfo {pages} {053001} (\bibinfo {year} {2016})}\BibitemShut {NoStop}%
\bibitem [{\citenamefont {Weedbrook}\ \emph {et~al.}(2012)\citenamefont {Weedbrook}, \citenamefont {Pirandola}, \citenamefont {Garc{\'\i}a-Patr{\'o}n}, \citenamefont {Cerf}, \citenamefont {Ralph}, \citenamefont {Shapiro},\ and\ \citenamefont {Lloyd}}]{GaussianQuantumInfo}%
  \BibitemOpen
  \bibfield  {author} {\bibinfo {author} {\bibfnamefont {C.}~\bibnamefont {Weedbrook}}, \bibinfo {author} {\bibfnamefont {S.}~\bibnamefont {Pirandola}}, \bibinfo {author} {\bibfnamefont {R.}~\bibnamefont {Garc{\'\i}a-Patr{\'o}n}}, \bibinfo {author} {\bibfnamefont {N.~J.}\ \bibnamefont {Cerf}}, \bibinfo {author} {\bibfnamefont {T.~C.}\ \bibnamefont {Ralph}}, \bibinfo {author} {\bibfnamefont {J.~H.}\ \bibnamefont {Shapiro}},\ and\ \bibinfo {author} {\bibfnamefont {S.}~\bibnamefont {Lloyd}},\ }\bibfield  {title} {\bibinfo {title} {Gaussian quantum information},\ }\href@noop {} {\bibfield  {journal} {\bibinfo  {journal} {Reviews of Modern Physics}\ }\textbf {\bibinfo {volume} {84}},\ \bibinfo {pages} {621} (\bibinfo {year} {2012})}\BibitemShut {NoStop}%
\bibitem [{\citenamefont {Aghaee~Rad}\ \emph {et~al.}(2025)\citenamefont {Aghaee~Rad}, \citenamefont {Ainsworth}, \citenamefont {Alexander}, \citenamefont {Altieri}, \citenamefont {Askarani}, \citenamefont {Baby}, \citenamefont {Banchi}, \citenamefont {Baragiola}, \citenamefont {Bourassa}, \citenamefont {Chadwick} \emph {et~al.}}]{aghaee2025scaling}%
  \BibitemOpen
  \bibfield  {author} {\bibinfo {author} {\bibfnamefont {H.}~\bibnamefont {Aghaee~Rad}}, \bibinfo {author} {\bibfnamefont {T.}~\bibnamefont {Ainsworth}}, \bibinfo {author} {\bibfnamefont {R.}~\bibnamefont {Alexander}}, \bibinfo {author} {\bibfnamefont {B.}~\bibnamefont {Altieri}}, \bibinfo {author} {\bibfnamefont {M.}~\bibnamefont {Askarani}}, \bibinfo {author} {\bibfnamefont {R.}~\bibnamefont {Baby}}, \bibinfo {author} {\bibfnamefont {L.}~\bibnamefont {Banchi}}, \bibinfo {author} {\bibfnamefont {B.}~\bibnamefont {Baragiola}}, \bibinfo {author} {\bibfnamefont {J.}~\bibnamefont {Bourassa}}, \bibinfo {author} {\bibfnamefont {R.}~\bibnamefont {Chadwick}}, \emph {et~al.},\ }\bibfield  {title} {\bibinfo {title} {Scaling and networking a modular photonic quantum computer},\ }\href@noop {} {\bibfield  {journal} {\bibinfo  {journal} {Nature}\ ,\ \bibinfo {pages} {1}} (\bibinfo {year} {2025})}\BibitemShut {NoStop}%
\bibitem [{psi(2025)}]{psiquantum2025manufacturable}%
  \BibitemOpen
  \bibfield  {title} {\bibinfo {title} {A manufacturable platform for photonic quantum computing},\ }\href@noop {} {\bibfield  {journal} {\bibinfo  {journal} {Nature}\ ,\ \bibinfo {pages} {1}} (\bibinfo {year} {2025})}\BibitemShut {NoStop}%
\bibitem [{\citenamefont {Caspani}\ \emph {et~al.}(2017)\citenamefont {Caspani}, \citenamefont {Xiong}, \citenamefont {Eggleton}, \citenamefont {Bajoni}, \citenamefont {Liscidini}, \citenamefont {Galli}, \citenamefont {Morandotti},\ and\ \citenamefont {Moss}}]{caspani2017integrated}%
  \BibitemOpen
  \bibfield  {author} {\bibinfo {author} {\bibfnamefont {L.}~\bibnamefont {Caspani}}, \bibinfo {author} {\bibfnamefont {C.}~\bibnamefont {Xiong}}, \bibinfo {author} {\bibfnamefont {B.~J.}\ \bibnamefont {Eggleton}}, \bibinfo {author} {\bibfnamefont {D.}~\bibnamefont {Bajoni}}, \bibinfo {author} {\bibfnamefont {M.}~\bibnamefont {Liscidini}}, \bibinfo {author} {\bibfnamefont {M.}~\bibnamefont {Galli}}, \bibinfo {author} {\bibfnamefont {R.}~\bibnamefont {Morandotti}},\ and\ \bibinfo {author} {\bibfnamefont {D.~J.}\ \bibnamefont {Moss}},\ }\bibfield  {title} {\bibinfo {title} {Integrated sources of photon quantum states based on nonlinear optics},\ }\href@noop {} {\bibfield  {journal} {\bibinfo  {journal} {Light: Science \& Applications}\ }\textbf {\bibinfo {volume} {6}},\ \bibinfo {pages} {e17100} (\bibinfo {year} {2017})}\BibitemShut {NoStop}%
\bibitem [{\citenamefont {Zhang}\ \emph {et~al.}(2021)\citenamefont {Zhang}, \citenamefont {Menotti}, \citenamefont {Tan}, \citenamefont {Vaidya}, \citenamefont {Mahler}, \citenamefont {Helt}, \citenamefont {Zatti}, \citenamefont {Liscidini}, \citenamefont {Morrison},\ and\ \citenamefont {Vernon}}]{zhang2021squeezed}%
  \BibitemOpen
  \bibfield  {author} {\bibinfo {author} {\bibfnamefont {Y.}~\bibnamefont {Zhang}}, \bibinfo {author} {\bibfnamefont {M.}~\bibnamefont {Menotti}}, \bibinfo {author} {\bibfnamefont {K.}~\bibnamefont {Tan}}, \bibinfo {author} {\bibfnamefont {V.}~\bibnamefont {Vaidya}}, \bibinfo {author} {\bibfnamefont {D.}~\bibnamefont {Mahler}}, \bibinfo {author} {\bibfnamefont {L.}~\bibnamefont {Helt}}, \bibinfo {author} {\bibfnamefont {L.}~\bibnamefont {Zatti}}, \bibinfo {author} {\bibfnamefont {M.}~\bibnamefont {Liscidini}}, \bibinfo {author} {\bibfnamefont {B.}~\bibnamefont {Morrison}},\ and\ \bibinfo {author} {\bibfnamefont {Z.}~\bibnamefont {Vernon}},\ }\bibfield  {title} {\bibinfo {title} {Squeezed light from a nanophotonic molecule},\ }\href@noop {} {\bibfield  {journal} {\bibinfo  {journal} {Nature communications}\ }\textbf {\bibinfo {volume} {12}},\ \bibinfo {pages} {2233} (\bibinfo {year} {2021})}\BibitemShut {NoStop}%
\bibitem [{\citenamefont {Ulanov}\ \emph {et~al.}(2025)\citenamefont {Ulanov}, \citenamefont {Ruhnke}, \citenamefont {Wildi},\ and\ \citenamefont {Herr}}]{ulanov2025quadrature}%
  \BibitemOpen
  \bibfield  {author} {\bibinfo {author} {\bibfnamefont {A.~E.}\ \bibnamefont {Ulanov}}, \bibinfo {author} {\bibfnamefont {B.}~\bibnamefont {Ruhnke}}, \bibinfo {author} {\bibfnamefont {T.}~\bibnamefont {Wildi}},\ and\ \bibinfo {author} {\bibfnamefont {T.}~\bibnamefont {Herr}},\ }\bibfield  {title} {\bibinfo {title} {Quadrature squeezing in a nanophotonic microresonator},\ }\href@noop {} {\bibfield  {journal} {\bibinfo  {journal} {arXiv preprint arXiv:2502.17337}\ } (\bibinfo {year} {2025})}\BibitemShut {NoStop}%
\bibitem [{\citenamefont {Shen}\ \emph {et~al.}(2025{\natexlab{a}})\citenamefont {Shen}, \citenamefont {Hsieh}, \citenamefont {Sridhar}, \citenamefont {Feldman}, \citenamefont {Chang}, \citenamefont {Smith},\ and\ \citenamefont {Dutt}}]{shen2025strong}%
  \BibitemOpen
  \bibfield  {author} {\bibinfo {author} {\bibfnamefont {Y.}~\bibnamefont {Shen}}, \bibinfo {author} {\bibfnamefont {P.-Y.}\ \bibnamefont {Hsieh}}, \bibinfo {author} {\bibfnamefont {S.~K.}\ \bibnamefont {Sridhar}}, \bibinfo {author} {\bibfnamefont {S.}~\bibnamefont {Feldman}}, \bibinfo {author} {\bibfnamefont {Y.-C.}\ \bibnamefont {Chang}}, \bibinfo {author} {\bibfnamefont {T.~A.}\ \bibnamefont {Smith}},\ and\ \bibinfo {author} {\bibfnamefont {A.}~\bibnamefont {Dutt}},\ }\bibfield  {title} {\bibinfo {title} {Strong nanophotonic quantum squeezing exceeding 3.5 db in a foundry-compatible kerr microresonator},\ }\href@noop {} {\bibfield  {journal} {\bibinfo  {journal} {Optica}\ }\textbf {\bibinfo {volume} {12}},\ \bibinfo {pages} {302} (\bibinfo {year} {2025}{\natexlab{a}})}\BibitemShut {NoStop}%
\bibitem [{\citenamefont {Shen}\ \emph {et~al.}(2025{\natexlab{b}})\citenamefont {Shen}, \citenamefont {Hsieh}, \citenamefont {Srinivasan}, \citenamefont {Henry}, \citenamefont {Moille}, \citenamefont {Sridhar}, \citenamefont {Restelli}, \citenamefont {Chang}, \citenamefont {Srinivasan}, \citenamefont {Smith} \emph {et~al.}}]{shen2025highly}%
  \BibitemOpen
  \bibfield  {author} {\bibinfo {author} {\bibfnamefont {Y.}~\bibnamefont {Shen}}, \bibinfo {author} {\bibfnamefont {P.-Y.}\ \bibnamefont {Hsieh}}, \bibinfo {author} {\bibfnamefont {D.}~\bibnamefont {Srinivasan}}, \bibinfo {author} {\bibfnamefont {A.}~\bibnamefont {Henry}}, \bibinfo {author} {\bibfnamefont {G.}~\bibnamefont {Moille}}, \bibinfo {author} {\bibfnamefont {S.~K.}\ \bibnamefont {Sridhar}}, \bibinfo {author} {\bibfnamefont {A.}~\bibnamefont {Restelli}}, \bibinfo {author} {\bibfnamefont {Y.-C.}\ \bibnamefont {Chang}}, \bibinfo {author} {\bibfnamefont {K.}~\bibnamefont {Srinivasan}}, \bibinfo {author} {\bibfnamefont {T.~A.}\ \bibnamefont {Smith}}, \emph {et~al.},\ }\bibfield  {title} {\bibinfo {title} {Highly squeezed nanophotonic quantum microcombs with broadband frequency tunability},\ }\href@noop {} {\bibfield  {journal} {\bibinfo  {journal} {arXiv preprint arXiv:2505.03734}\ } (\bibinfo {year} {2025}{\natexlab{b}})}\BibitemShut {NoStop}%
\bibitem [{\citenamefont {Larsen}\ \emph {et~al.}(2025)\citenamefont {Larsen}, \citenamefont {Bourassa}, \citenamefont {Kocsis}, \citenamefont {Tasker}, \citenamefont {Chadwick}, \citenamefont {Gonz{\'a}lez-Arciniegas}, \citenamefont {Hastrup}, \citenamefont {Lopetegui-Gonz{\'a}lez}, \citenamefont {Miatto}, \citenamefont {Motamedi} \emph {et~al.}}]{larsen2025integrated}%
  \BibitemOpen
  \bibfield  {author} {\bibinfo {author} {\bibfnamefont {M.}~\bibnamefont {Larsen}}, \bibinfo {author} {\bibfnamefont {J.}~\bibnamefont {Bourassa}}, \bibinfo {author} {\bibfnamefont {S.}~\bibnamefont {Kocsis}}, \bibinfo {author} {\bibfnamefont {J.}~\bibnamefont {Tasker}}, \bibinfo {author} {\bibfnamefont {R.}~\bibnamefont {Chadwick}}, \bibinfo {author} {\bibfnamefont {C.}~\bibnamefont {Gonz{\'a}lez-Arciniegas}}, \bibinfo {author} {\bibfnamefont {J.}~\bibnamefont {Hastrup}}, \bibinfo {author} {\bibfnamefont {C.}~\bibnamefont {Lopetegui-Gonz{\'a}lez}}, \bibinfo {author} {\bibfnamefont {F.}~\bibnamefont {Miatto}}, \bibinfo {author} {\bibfnamefont {A.}~\bibnamefont {Motamedi}}, \emph {et~al.},\ }\bibfield  {title} {\bibinfo {title} {Integrated photonic source of gottesman--kitaev--preskill qubits},\ }\href@noop {} {\bibfield  {journal} {\bibinfo  {journal} {Nature}\ ,\ \bibinfo {pages} {1}} (\bibinfo {year} {2025})}\BibitemShut {NoStop}%
\bibitem [{\citenamefont {Jahanbozorgi}\ \emph {et~al.}(2023)\citenamefont {Jahanbozorgi}, \citenamefont {Yang}, \citenamefont {Sun}, \citenamefont {Chen}, \citenamefont {Liu}, \citenamefont {Wang},\ and\ \citenamefont {Yi}}]{jahanbozorgi2023generation}%
  \BibitemOpen
  \bibfield  {author} {\bibinfo {author} {\bibfnamefont {M.}~\bibnamefont {Jahanbozorgi}}, \bibinfo {author} {\bibfnamefont {Z.}~\bibnamefont {Yang}}, \bibinfo {author} {\bibfnamefont {S.}~\bibnamefont {Sun}}, \bibinfo {author} {\bibfnamefont {H.}~\bibnamefont {Chen}}, \bibinfo {author} {\bibfnamefont {R.}~\bibnamefont {Liu}}, \bibinfo {author} {\bibfnamefont {B.}~\bibnamefont {Wang}},\ and\ \bibinfo {author} {\bibfnamefont {X.}~\bibnamefont {Yi}},\ }\bibfield  {title} {\bibinfo {title} {Generation of squeezed quantum microcombs with silicon nitride integrated photonic circuits},\ }\href@noop {} {\bibfield  {journal} {\bibinfo  {journal} {Optica}\ }\textbf {\bibinfo {volume} {10}},\ \bibinfo {pages} {1100} (\bibinfo {year} {2023})}\BibitemShut {NoStop}%
\bibitem [{\citenamefont {Jia}\ \emph {et~al.}(2025)\citenamefont {Jia}, \citenamefont {Zhai}, \citenamefont {Zhu}, \citenamefont {You}, \citenamefont {Cao}, \citenamefont {Zhang}, \citenamefont {Zheng}, \citenamefont {Fu}, \citenamefont {Mao}, \citenamefont {Dai} \emph {et~al.}}]{jia2025continuous}%
  \BibitemOpen
  \bibfield  {author} {\bibinfo {author} {\bibfnamefont {X.}~\bibnamefont {Jia}}, \bibinfo {author} {\bibfnamefont {C.}~\bibnamefont {Zhai}}, \bibinfo {author} {\bibfnamefont {X.}~\bibnamefont {Zhu}}, \bibinfo {author} {\bibfnamefont {C.}~\bibnamefont {You}}, \bibinfo {author} {\bibfnamefont {Y.}~\bibnamefont {Cao}}, \bibinfo {author} {\bibfnamefont {X.}~\bibnamefont {Zhang}}, \bibinfo {author} {\bibfnamefont {Y.}~\bibnamefont {Zheng}}, \bibinfo {author} {\bibfnamefont {Z.}~\bibnamefont {Fu}}, \bibinfo {author} {\bibfnamefont {J.}~\bibnamefont {Mao}}, \bibinfo {author} {\bibfnamefont {T.}~\bibnamefont {Dai}}, \emph {et~al.},\ }\bibfield  {title} {\bibinfo {title} {Continuous-variable multipartite entanglement in an integrated microcomb},\ }\href@noop {} {\bibfield  {journal} {\bibinfo  {journal} {Nature}\ ,\ \bibinfo {pages} {1}} (\bibinfo {year} {2025})}\BibitemShut {NoStop}%
\bibitem [{\citenamefont {Gouzien}\ \emph {et~al.}(2023)\citenamefont {Gouzien}, \citenamefont {Labont{\'e}}, \citenamefont {Etesse}, \citenamefont {Zavatta}, \citenamefont {Tanzilli}, \citenamefont {d'Auria},\ and\ \citenamefont {Patera}}]{gouzien2023hidden}%
  \BibitemOpen
  \bibfield  {author} {\bibinfo {author} {\bibfnamefont {{\'E}.}~\bibnamefont {Gouzien}}, \bibinfo {author} {\bibfnamefont {L.}~\bibnamefont {Labont{\'e}}}, \bibinfo {author} {\bibfnamefont {J.}~\bibnamefont {Etesse}}, \bibinfo {author} {\bibfnamefont {A.}~\bibnamefont {Zavatta}}, \bibinfo {author} {\bibfnamefont {S.}~\bibnamefont {Tanzilli}}, \bibinfo {author} {\bibfnamefont {V.}~\bibnamefont {d'Auria}},\ and\ \bibinfo {author} {\bibfnamefont {G.}~\bibnamefont {Patera}},\ }\bibfield  {title} {\bibinfo {title} {Hidden and detectable squeezing from microresonators},\ }\href@noop {} {\bibfield  {journal} {\bibinfo  {journal} {Physical Review Research}\ }\textbf {\bibinfo {volume} {5}},\ \bibinfo {pages} {023178} (\bibinfo {year} {2023})}\BibitemShut {NoStop}%
\bibitem [{\citenamefont {Wang}\ \emph {et~al.}(2025)\citenamefont {Wang}, \citenamefont {Li}, \citenamefont {Wang}, \citenamefont {Zhou}, \citenamefont {Cheng}, \citenamefont {Jing}, \citenamefont {Sun}, \citenamefont {Li}, \citenamefont {Li}, \citenamefont {Wu} \emph {et~al.}}]{wang2025large}%
  \BibitemOpen
  \bibfield  {author} {\bibinfo {author} {\bibfnamefont {Z.}~\bibnamefont {Wang}}, \bibinfo {author} {\bibfnamefont {K.}~\bibnamefont {Li}}, \bibinfo {author} {\bibfnamefont {Y.}~\bibnamefont {Wang}}, \bibinfo {author} {\bibfnamefont {X.}~\bibnamefont {Zhou}}, \bibinfo {author} {\bibfnamefont {Y.}~\bibnamefont {Cheng}}, \bibinfo {author} {\bibfnamefont {B.}~\bibnamefont {Jing}}, \bibinfo {author} {\bibfnamefont {F.}~\bibnamefont {Sun}}, \bibinfo {author} {\bibfnamefont {J.}~\bibnamefont {Li}}, \bibinfo {author} {\bibfnamefont {Z.}~\bibnamefont {Li}}, \bibinfo {author} {\bibfnamefont {B.}~\bibnamefont {Wu}}, \emph {et~al.},\ }\bibfield  {title} {\bibinfo {title} {Large-scale cluster quantum microcombs},\ }\href@noop {} {\bibfield  {journal} {\bibinfo  {journal} {Light: Science \& Applications}\ }\textbf {\bibinfo {volume} {14}},\ \bibinfo {pages} {164} (\bibinfo {year} {2025})}\BibitemShut {NoStop}%
\bibitem [{\citenamefont {Quesada}\ \emph {et~al.}(2022)\citenamefont {Quesada}, \citenamefont {Helt}, \citenamefont {Menotti}, \citenamefont {Liscidini},\ and\ \citenamefont {Sipe}}]{BEYOND}%
  \BibitemOpen
  \bibfield  {author} {\bibinfo {author} {\bibfnamefont {N.}~\bibnamefont {Quesada}}, \bibinfo {author} {\bibfnamefont {L.}~\bibnamefont {Helt}}, \bibinfo {author} {\bibfnamefont {M.}~\bibnamefont {Menotti}}, \bibinfo {author} {\bibfnamefont {M.}~\bibnamefont {Liscidini}},\ and\ \bibinfo {author} {\bibfnamefont {J.}~\bibnamefont {Sipe}},\ }\bibfield  {title} {\bibinfo {title} {Beyond photon pairs—nonlinear quantum photonics in the high-gain regime: a tutorial},\ }\href {https://doi.org/10.1364/AOP.445496} {\bibfield  {journal} {\bibinfo  {journal} {Advances in Optics and Photonics}\ }\textbf {\bibinfo {volume} {14}},\ \bibinfo {pages} {291} (\bibinfo {year} {2022})}\BibitemShut {NoStop}%
\bibitem [{\citenamefont {Sloan}\ \emph {et~al.}(2025)\citenamefont {Sloan}, \citenamefont {Viola}, \citenamefont {Liscidini},\ and\ \citenamefont {Sipe}}]{sloan2025high}%
  \BibitemOpen
  \bibfield  {author} {\bibinfo {author} {\bibfnamefont {M.}~\bibnamefont {Sloan}}, \bibinfo {author} {\bibfnamefont {A.}~\bibnamefont {Viola}}, \bibinfo {author} {\bibfnamefont {M.}~\bibnamefont {Liscidini}},\ and\ \bibinfo {author} {\bibfnamefont {J.}~\bibnamefont {Sipe}},\ }\bibfield  {title} {\bibinfo {title} {High-gain squeezing in lossy resonators: An asymptotic-field approach},\ }\href@noop {} {\bibfield  {journal} {\bibinfo  {journal} {Physical Review A}\ }\textbf {\bibinfo {volume} {111}},\ \bibinfo {pages} {063502} (\bibinfo {year} {2025})}\BibitemShut {NoStop}%
\bibitem [{\citenamefont {Vendromin}\ \emph {et~al.}(2024)\citenamefont {Vendromin}, \citenamefont {Liu}, \citenamefont {Yang},\ and\ \citenamefont {Sipe}}]{vendromin2024highly}%
  \BibitemOpen
  \bibfield  {author} {\bibinfo {author} {\bibfnamefont {C.}~\bibnamefont {Vendromin}}, \bibinfo {author} {\bibfnamefont {Y.}~\bibnamefont {Liu}}, \bibinfo {author} {\bibfnamefont {Z.}~\bibnamefont {Yang}},\ and\ \bibinfo {author} {\bibfnamefont {J.~E.}\ \bibnamefont {Sipe}},\ }\bibfield  {title} {\bibinfo {title} {Highly squeezed states in ring resonators: Beyond the undepleted-pump approximation},\ }\href@noop {} {\bibfield  {journal} {\bibinfo  {journal} {Physical Review A}\ }\textbf {\bibinfo {volume} {110}},\ \bibinfo {pages} {033709} (\bibinfo {year} {2024})}\BibitemShut {NoStop}%
\bibitem [{\citenamefont {Vernon}\ and\ \citenamefont {Sipe}(2015)}]{Strongly_Zachary}%
  \BibitemOpen
  \bibfield  {author} {\bibinfo {author} {\bibfnamefont {Z.}~\bibnamefont {Vernon}}\ and\ \bibinfo {author} {\bibfnamefont {J.~E.}\ \bibnamefont {Sipe}},\ }\bibfield  {title} {\bibinfo {title} {Strongly driven nonlinear quantum optics in microring resonators},\ }\href {https://doi.org/10.1103/PhysRevA.92.033840} {\bibfield  {journal} {\bibinfo  {journal} {Phys. Rev. A}\ }\textbf {\bibinfo {volume} {92}},\ \bibinfo {pages} {033840} (\bibinfo {year} {2015})}\BibitemShut {NoStop}%
\bibitem [{\citenamefont {Vernon}\ \emph {et~al.}(2019)\citenamefont {Vernon}, \citenamefont {Quesada}, \citenamefont {Liscidini}, \citenamefont {Morrison}, \citenamefont {Menotti}, \citenamefont {Tan},\ and\ \citenamefont {Sipe}}]{vernon2019scalable}%
  \BibitemOpen
  \bibfield  {author} {\bibinfo {author} {\bibfnamefont {Z.}~\bibnamefont {Vernon}}, \bibinfo {author} {\bibfnamefont {N.}~\bibnamefont {Quesada}}, \bibinfo {author} {\bibfnamefont {M.}~\bibnamefont {Liscidini}}, \bibinfo {author} {\bibfnamefont {B.}~\bibnamefont {Morrison}}, \bibinfo {author} {\bibfnamefont {M.}~\bibnamefont {Menotti}}, \bibinfo {author} {\bibfnamefont {K.}~\bibnamefont {Tan}},\ and\ \bibinfo {author} {\bibfnamefont {J.}~\bibnamefont {Sipe}},\ }\bibfield  {title} {\bibinfo {title} {Scalable squeezed-light source for continuous-variable quantum sampling},\ }\href@noop {} {\bibfield  {journal} {\bibinfo  {journal} {Physical Review Applied}\ }\textbf {\bibinfo {volume} {12}},\ \bibinfo {pages} {064024} (\bibinfo {year} {2019})}\BibitemShut {NoStop}%
\bibitem [{\citenamefont {Guidry}\ \emph {et~al.}(2022)\citenamefont {Guidry}, \citenamefont {Lukin}, \citenamefont {Yang}, \citenamefont {Trivedi},\ and\ \citenamefont {Vu{\v{c}}kovi{\'c}}}]{guidry2022quantum}%
  \BibitemOpen
  \bibfield  {author} {\bibinfo {author} {\bibfnamefont {M.~A.}\ \bibnamefont {Guidry}}, \bibinfo {author} {\bibfnamefont {D.~M.}\ \bibnamefont {Lukin}}, \bibinfo {author} {\bibfnamefont {K.~Y.}\ \bibnamefont {Yang}}, \bibinfo {author} {\bibfnamefont {R.}~\bibnamefont {Trivedi}},\ and\ \bibinfo {author} {\bibfnamefont {J.}~\bibnamefont {Vu{\v{c}}kovi{\'c}}},\ }\bibfield  {title} {\bibinfo {title} {Quantum optics of soliton microcombs},\ }\href@noop {} {\bibfield  {journal} {\bibinfo  {journal} {Nature Photonics}\ }\textbf {\bibinfo {volume} {16}},\ \bibinfo {pages} {52} (\bibinfo {year} {2022})}\BibitemShut {NoStop}%
\bibitem [{\citenamefont {Ramelow}\ \emph {et~al.}(2019)\citenamefont {Ramelow}, \citenamefont {Farsi}, \citenamefont {Vernon}, \citenamefont {Clemmen}, \citenamefont {Ji}, \citenamefont {Sipe}, \citenamefont {Liscidini}, \citenamefont {Lipson},\ and\ \citenamefont {Gaeta}}]{ramelow2019strong}%
  \BibitemOpen
  \bibfield  {author} {\bibinfo {author} {\bibfnamefont {S.}~\bibnamefont {Ramelow}}, \bibinfo {author} {\bibfnamefont {A.}~\bibnamefont {Farsi}}, \bibinfo {author} {\bibfnamefont {Z.}~\bibnamefont {Vernon}}, \bibinfo {author} {\bibfnamefont {S.}~\bibnamefont {Clemmen}}, \bibinfo {author} {\bibfnamefont {X.}~\bibnamefont {Ji}}, \bibinfo {author} {\bibfnamefont {J.}~\bibnamefont {Sipe}}, \bibinfo {author} {\bibfnamefont {M.}~\bibnamefont {Liscidini}}, \bibinfo {author} {\bibfnamefont {M.}~\bibnamefont {Lipson}},\ and\ \bibinfo {author} {\bibfnamefont {A.~L.}\ \bibnamefont {Gaeta}},\ }\bibfield  {title} {\bibinfo {title} {Strong nonlinear coupling in a si 3 n 4 ring resonator},\ }\href@noop {} {\bibfield  {journal} {\bibinfo  {journal} {Physical review letters}\ }\textbf {\bibinfo {volume} {122}},\ \bibinfo {pages} {153906} (\bibinfo {year} {2019})}\BibitemShut {NoStop}%
\bibitem [{\citenamefont {Kim}\ \emph {et~al.}(2025)\citenamefont {Kim}, \citenamefont {Jeon},\ and\ \citenamefont {Sohn}}]{kim2025simulating}%
  \BibitemOpen
  \bibfield  {author} {\bibinfo {author} {\bibfnamefont {Y.}~\bibnamefont {Kim}}, \bibinfo {author} {\bibfnamefont {S.}~\bibnamefont {Jeon}},\ and\ \bibinfo {author} {\bibfnamefont {Y.-I.}\ \bibnamefont {Sohn}},\ }\bibfield  {title} {\bibinfo {title} {Simulating quantum light in lossy microring resonators driven by strong pulses},\ }\href@noop {} {\bibfield  {journal} {\bibinfo  {journal} {Physical Review Applied}\ }\textbf {\bibinfo {volume} {23}},\ \bibinfo {pages} {054045} (\bibinfo {year} {2025})}\BibitemShut {NoStop}%
\bibitem [{\citenamefont {Cui}\ \emph {et~al.}(2021)\citenamefont {Cui}, \citenamefont {Gagatsos}, \citenamefont {Guha},\ and\ \citenamefont {Fan}}]{cui2021high}%
  \BibitemOpen
  \bibfield  {author} {\bibinfo {author} {\bibfnamefont {C.}~\bibnamefont {Cui}}, \bibinfo {author} {\bibfnamefont {C.~N.}\ \bibnamefont {Gagatsos}}, \bibinfo {author} {\bibfnamefont {S.}~\bibnamefont {Guha}},\ and\ \bibinfo {author} {\bibfnamefont {L.}~\bibnamefont {Fan}},\ }\bibfield  {title} {\bibinfo {title} {High-purity pulsed squeezing generation with integrated photonics},\ }\href@noop {} {\bibfield  {journal} {\bibinfo  {journal} {Physical Review Research}\ }\textbf {\bibinfo {volume} {3}},\ \bibinfo {pages} {013199} (\bibinfo {year} {2021})}\BibitemShut {NoStop}%
\bibitem [{\citenamefont {Arrazola}\ \emph {et~al.}(2021)\citenamefont {Arrazola}, \citenamefont {Bergholm}, \citenamefont {Br{\'a}dler}, \citenamefont {Bromley}, \citenamefont {Collins}, \citenamefont {Dhand}, \citenamefont {Fumagalli}, \citenamefont {Gerrits}, \citenamefont {Goussev}, \citenamefont {Helt} \emph {et~al.}}]{arrazola2021quantum}%
  \BibitemOpen
  \bibfield  {author} {\bibinfo {author} {\bibfnamefont {J.~M.}\ \bibnamefont {Arrazola}}, \bibinfo {author} {\bibfnamefont {V.}~\bibnamefont {Bergholm}}, \bibinfo {author} {\bibfnamefont {K.}~\bibnamefont {Br{\'a}dler}}, \bibinfo {author} {\bibfnamefont {T.~R.}\ \bibnamefont {Bromley}}, \bibinfo {author} {\bibfnamefont {M.~J.}\ \bibnamefont {Collins}}, \bibinfo {author} {\bibfnamefont {I.}~\bibnamefont {Dhand}}, \bibinfo {author} {\bibfnamefont {A.}~\bibnamefont {Fumagalli}}, \bibinfo {author} {\bibfnamefont {T.}~\bibnamefont {Gerrits}}, \bibinfo {author} {\bibfnamefont {A.}~\bibnamefont {Goussev}}, \bibinfo {author} {\bibfnamefont {L.~G.}\ \bibnamefont {Helt}}, \emph {et~al.},\ }\bibfield  {title} {\bibinfo {title} {Quantum circuits with many photons on a programmable nanophotonic chip},\ }\href@noop {} {\bibfield  {journal} {\bibinfo  {journal} {Nature}\ }\textbf {\bibinfo {volume} {591}},\ \bibinfo {pages} {54} (\bibinfo {year} {2021})}\BibitemShut {NoStop}%
\bibitem [{\citenamefont {Dignam}\ and\ \citenamefont {Liscidini}(2025)}]{f2y1-scgw}%
  \BibitemOpen
  \bibfield  {author} {\bibinfo {author} {\bibfnamefont {M.~M.}\ \bibnamefont {Dignam}}\ and\ \bibinfo {author} {\bibfnamefont {M.}~\bibnamefont {Liscidini}},\ }\bibfield  {title} {\bibinfo {title} {Modeling and optimization of pulsed-squeezed-state generation in a ring-resonator system},\ }\href {https://doi.org/10.1103/f2y1-scgw} {\bibfield  {journal} {\bibinfo  {journal} {Phys. Rev. A}\ }\textbf {\bibinfo {volume} {112}},\ \bibinfo {pages} {013710} (\bibinfo {year} {2025})}\BibitemShut {NoStop}%
\bibitem [{\citenamefont {Banic}\ \emph {et~al.}(2022)\citenamefont {Banic}, \citenamefont {Zatti}, \citenamefont {Liscidini},\ and\ \citenamefont {Sipe}}]{banic2022two}%
  \BibitemOpen
  \bibfield  {author} {\bibinfo {author} {\bibfnamefont {M.}~\bibnamefont {Banic}}, \bibinfo {author} {\bibfnamefont {L.}~\bibnamefont {Zatti}}, \bibinfo {author} {\bibfnamefont {M.}~\bibnamefont {Liscidini}},\ and\ \bibinfo {author} {\bibfnamefont {J.}~\bibnamefont {Sipe}},\ }\bibfield  {title} {\bibinfo {title} {Two strategies for modeling nonlinear optics in lossy integrated photonic structures},\ }\href@noop {} {\bibfield  {journal} {\bibinfo  {journal} {Physical Review A}\ }\textbf {\bibinfo {volume} {106}},\ \bibinfo {pages} {043707} (\bibinfo {year} {2022})}\BibitemShut {NoStop}%
\bibitem [{\citenamefont {Brusaschi}\ \emph {et~al.}(2024)\citenamefont {Brusaschi}, \citenamefont {Borghi}, \citenamefont {Bacchi}, \citenamefont {Liscidini}, \citenamefont {Galli},\ and\ \citenamefont {Bajoni}}]{Brusaschi:24}%
  \BibitemOpen
  \bibfield  {author} {\bibinfo {author} {\bibfnamefont {E.}~\bibnamefont {Brusaschi}}, \bibinfo {author} {\bibfnamefont {M.}~\bibnamefont {Borghi}}, \bibinfo {author} {\bibfnamefont {M.}~\bibnamefont {Bacchi}}, \bibinfo {author} {\bibfnamefont {M.}~\bibnamefont {Liscidini}}, \bibinfo {author} {\bibfnamefont {M.}~\bibnamefont {Galli}},\ and\ \bibinfo {author} {\bibfnamefont {D.}~\bibnamefont {Bajoni}},\ }\bibfield  {title} {\bibinfo {title} {Photon number distribution of squeezed light from a silicon nitride microresonator measured without photon number resolving detectors},\ }\href {https://doi.org/10.1364/OPTICAQ.528566} {\bibfield  {journal} {\bibinfo  {journal} {Optica Quantum}\ }\textbf {\bibinfo {volume} {2}},\ \bibinfo {pages} {214} (\bibinfo {year} {2024})}\BibitemShut {NoStop}%
\bibitem [{\citenamefont {Johansson}\ \emph {et~al.}(2012)\citenamefont {Johansson}, \citenamefont {Nation},\ and\ \citenamefont {Nori}}]{QuTiP}%
  \BibitemOpen
  \bibfield  {author} {\bibinfo {author} {\bibfnamefont {J.}~\bibnamefont {Johansson}}, \bibinfo {author} {\bibfnamefont {P.}~\bibnamefont {Nation}},\ and\ \bibinfo {author} {\bibfnamefont {F.}~\bibnamefont {Nori}},\ }\bibfield  {title} {\bibinfo {title} {Qutip: An open-source python framework for the dynamics of open quantum systems},\ }\href {https://doi.org/https://doi.org/10.1016/j.cpc.2012.02.021} {\bibfield  {journal} {\bibinfo  {journal} {Computer Physics Communications}\ }\textbf {\bibinfo {volume} {183}},\ \bibinfo {pages} {1760} (\bibinfo {year} {2012})}\BibitemShut {NoStop}%
\bibitem [{\citenamefont {Stone}\ \emph {et~al.}(2025)\citenamefont {Stone}, \citenamefont {Rukh}, \citenamefont {Colaci{\'o}n},\ and\ \citenamefont {Drake}}]{stone2025reduction}%
  \BibitemOpen
  \bibfield  {author} {\bibinfo {author} {\bibfnamefont {B.~D.}\ \bibnamefont {Stone}}, \bibinfo {author} {\bibfnamefont {L.}~\bibnamefont {Rukh}}, \bibinfo {author} {\bibfnamefont {G.~M.}\ \bibnamefont {Colaci{\'o}n}},\ and\ \bibinfo {author} {\bibfnamefont {T.~E.}\ \bibnamefont {Drake}},\ }\bibfield  {title} {\bibinfo {title} {Reduction of thermal instability of soliton states in coupled kerr-microresonators},\ }\href@noop {} {\bibfield  {journal} {\bibinfo  {journal} {APL Photonics}\ }\textbf {\bibinfo {volume} {10}} (\bibinfo {year} {2025})}\BibitemShut {NoStop}%
\bibitem [{\citenamefont {Christ}\ \emph {et~al.}(2011)\citenamefont {Christ}, \citenamefont {Laiho}, \citenamefont {Eckstein}, \citenamefont {Cassemiro},\ and\ \citenamefont {Silberhorn}}]{christ2011probing}%
  \BibitemOpen
  \bibfield  {author} {\bibinfo {author} {\bibfnamefont {A.}~\bibnamefont {Christ}}, \bibinfo {author} {\bibfnamefont {K.}~\bibnamefont {Laiho}}, \bibinfo {author} {\bibfnamefont {A.}~\bibnamefont {Eckstein}}, \bibinfo {author} {\bibfnamefont {K.~N.}\ \bibnamefont {Cassemiro}},\ and\ \bibinfo {author} {\bibfnamefont {C.}~\bibnamefont {Silberhorn}},\ }\bibfield  {title} {\bibinfo {title} {Probing multimode squeezing with correlation functions},\ }\href@noop {} {\bibfield  {journal} {\bibinfo  {journal} {New Journal of Physics}\ }\textbf {\bibinfo {volume} {13}},\ \bibinfo {pages} {033027} (\bibinfo {year} {2011})}\BibitemShut {NoStop}%
\bibitem [{\citenamefont {Laiho}\ \emph {et~al.}(2022)\citenamefont {Laiho}, \citenamefont {Dirmeier}, \citenamefont {Schmidt}, \citenamefont {Reitzenstein},\ and\ \citenamefont {Marquardt}}]{laiho2022measuring}%
  \BibitemOpen
  \bibfield  {author} {\bibinfo {author} {\bibfnamefont {K.}~\bibnamefont {Laiho}}, \bibinfo {author} {\bibfnamefont {T.}~\bibnamefont {Dirmeier}}, \bibinfo {author} {\bibfnamefont {M.}~\bibnamefont {Schmidt}}, \bibinfo {author} {\bibfnamefont {S.}~\bibnamefont {Reitzenstein}},\ and\ \bibinfo {author} {\bibfnamefont {C.}~\bibnamefont {Marquardt}},\ }\bibfield  {title} {\bibinfo {title} {Measuring higher-order photon correlations of faint quantum light: a short review},\ }\href@noop {} {\bibfield  {journal} {\bibinfo  {journal} {Physics Letters A}\ }\textbf {\bibinfo {volume} {435}},\ \bibinfo {pages} {128059} (\bibinfo {year} {2022})}\BibitemShut {NoStop}%
\bibitem [{\citenamefont {Borghi}\ \emph {et~al.}(2024)\citenamefont {Borghi}, \citenamefont {Pagano}, \citenamefont {Liscidini}, \citenamefont {Bajoni},\ and\ \citenamefont {Galli}}]{borghi2024uncorrelated}%
  \BibitemOpen
  \bibfield  {author} {\bibinfo {author} {\bibfnamefont {M.}~\bibnamefont {Borghi}}, \bibinfo {author} {\bibfnamefont {P.~L.}\ \bibnamefont {Pagano}}, \bibinfo {author} {\bibfnamefont {M.}~\bibnamefont {Liscidini}}, \bibinfo {author} {\bibfnamefont {D.}~\bibnamefont {Bajoni}},\ and\ \bibinfo {author} {\bibfnamefont {M.}~\bibnamefont {Galli}},\ }\bibfield  {title} {\bibinfo {title} {Uncorrelated photon pair generation from an integrated silicon nitride resonator measured by time-resolved coincidence detection},\ }\href@noop {} {\bibfield  {journal} {\bibinfo  {journal} {Optics Letters}\ }\textbf {\bibinfo {volume} {49}},\ \bibinfo {pages} {3966} (\bibinfo {year} {2024})}\BibitemShut {NoStop}%
\bibitem [{\citenamefont {Christensen}\ \emph {et~al.}(2018)\citenamefont {Christensen}, \citenamefont {Koefoed}, \citenamefont {Rottwitt},\ and\ \citenamefont {McKinstrie}}]{christensen2018engineering}%
  \BibitemOpen
  \bibfield  {author} {\bibinfo {author} {\bibfnamefont {J.~B.}\ \bibnamefont {Christensen}}, \bibinfo {author} {\bibfnamefont {J.~G.}\ \bibnamefont {Koefoed}}, \bibinfo {author} {\bibfnamefont {K.}~\bibnamefont {Rottwitt}},\ and\ \bibinfo {author} {\bibfnamefont {C.}~\bibnamefont {McKinstrie}},\ }\bibfield  {title} {\bibinfo {title} {Engineering spectrally unentangled photon pairs from nonlinear microring resonators by pump manipulation},\ }\href@noop {} {\bibfield  {journal} {\bibinfo  {journal} {Optics letters}\ }\textbf {\bibinfo {volume} {43}},\ \bibinfo {pages} {859} (\bibinfo {year} {2018})}\BibitemShut {NoStop}%
\bibitem [{\citenamefont {Loudon}(2000)}]{loudon2000quantum}%
  \BibitemOpen
  \bibfield  {author} {\bibinfo {author} {\bibfnamefont {R.}~\bibnamefont {Loudon}},\ }\href@noop {} {\emph {\bibinfo {title} {The quantum theory of light}}}\ (\bibinfo  {publisher} {OUP Oxford},\ \bibinfo {year} {2000})\BibitemShut {NoStop}%
\bibitem [{\citenamefont {Christ}\ \emph {et~al.}(2013)\citenamefont {Christ}, \citenamefont {Brecht}, \citenamefont {Mauerer},\ and\ \citenamefont {Silberhorn}}]{christ2013theory}%
  \BibitemOpen
  \bibfield  {author} {\bibinfo {author} {\bibfnamefont {A.}~\bibnamefont {Christ}}, \bibinfo {author} {\bibfnamefont {B.}~\bibnamefont {Brecht}}, \bibinfo {author} {\bibfnamefont {W.}~\bibnamefont {Mauerer}},\ and\ \bibinfo {author} {\bibfnamefont {C.}~\bibnamefont {Silberhorn}},\ }\bibfield  {title} {\bibinfo {title} {Theory of quantum frequency conversion and type-ii parametric down-conversion in the high-gain regime},\ }\href@noop {} {\bibfield  {journal} {\bibinfo  {journal} {New Journal of Physics}\ }\textbf {\bibinfo {volume} {15}},\ \bibinfo {pages} {053038} (\bibinfo {year} {2013})}\BibitemShut {NoStop}%
\bibitem [{\citenamefont {Triginer}\ \emph {et~al.}(2020)\citenamefont {Triginer}, \citenamefont {Vidrighin}, \citenamefont {Quesada}, \citenamefont {Eckstein}, \citenamefont {Moore}, \citenamefont {Kolthammer}, \citenamefont {Sipe},\ and\ \citenamefont {Walmsley}}]{triginer2020understanding}%
  \BibitemOpen
  \bibfield  {author} {\bibinfo {author} {\bibfnamefont {G.}~\bibnamefont {Triginer}}, \bibinfo {author} {\bibfnamefont {M.~D.}\ \bibnamefont {Vidrighin}}, \bibinfo {author} {\bibfnamefont {N.}~\bibnamefont {Quesada}}, \bibinfo {author} {\bibfnamefont {A.}~\bibnamefont {Eckstein}}, \bibinfo {author} {\bibfnamefont {M.}~\bibnamefont {Moore}}, \bibinfo {author} {\bibfnamefont {W.~S.}\ \bibnamefont {Kolthammer}}, \bibinfo {author} {\bibfnamefont {J.}~\bibnamefont {Sipe}},\ and\ \bibinfo {author} {\bibfnamefont {I.~A.}\ \bibnamefont {Walmsley}},\ }\bibfield  {title} {\bibinfo {title} {Understanding high-gain twin-beam sources using cascaded stimulated emission},\ }\href@noop {} {\bibfield  {journal} {\bibinfo  {journal} {Physical Review X}\ }\textbf {\bibinfo {volume} {10}},\ \bibinfo {pages} {031063} (\bibinfo {year} {2020})}\BibitemShut {NoStop}%
\bibitem [{\citenamefont {Vernon}\ \emph {et~al.}(2017)\citenamefont {Vernon}, \citenamefont {Menotti}, \citenamefont {Tison}, \citenamefont {Steidle}, \citenamefont {Fanto}, \citenamefont {Thomas}, \citenamefont {Preble}, \citenamefont {Smith}, \citenamefont {Alsing}, \citenamefont {Liscidini} \emph {et~al.}}]{vernon2017truly}%
  \BibitemOpen
  \bibfield  {author} {\bibinfo {author} {\bibfnamefont {Z.}~\bibnamefont {Vernon}}, \bibinfo {author} {\bibfnamefont {M.}~\bibnamefont {Menotti}}, \bibinfo {author} {\bibfnamefont {C.}~\bibnamefont {Tison}}, \bibinfo {author} {\bibfnamefont {J.}~\bibnamefont {Steidle}}, \bibinfo {author} {\bibfnamefont {M.}~\bibnamefont {Fanto}}, \bibinfo {author} {\bibfnamefont {P.}~\bibnamefont {Thomas}}, \bibinfo {author} {\bibfnamefont {S.}~\bibnamefont {Preble}}, \bibinfo {author} {\bibfnamefont {A.}~\bibnamefont {Smith}}, \bibinfo {author} {\bibfnamefont {P.}~\bibnamefont {Alsing}}, \bibinfo {author} {\bibfnamefont {M.}~\bibnamefont {Liscidini}}, \emph {et~al.},\ }\bibfield  {title} {\bibinfo {title} {Truly unentangled photon pairs without spectral filtering},\ }\href@noop {} {\bibfield  {journal} {\bibinfo  {journal} {Optics letters}\ }\textbf {\bibinfo {volume} {42}},\ \bibinfo {pages} {3638} (\bibinfo {year} {2017})}\BibitemShut {NoStop}%
\bibitem [{\citenamefont {Sz{\'e}kely}\ \emph {et~al.}(2004)\citenamefont {Sz{\'e}kely}, \citenamefont {Rizzo} \emph {et~al.}}]{szekely2004testing}%
  \BibitemOpen
  \bibfield  {author} {\bibinfo {author} {\bibfnamefont {G.~J.}\ \bibnamefont {Sz{\'e}kely}}, \bibinfo {author} {\bibfnamefont {M.~L.}\ \bibnamefont {Rizzo}}, \emph {et~al.},\ }\bibfield  {title} {\bibinfo {title} {Testing for equal distributions in high dimension},\ }\href@noop {} {\bibfield  {journal} {\bibinfo  {journal} {InterStat}\ }\textbf {\bibinfo {volume} {5}},\ \bibinfo {pages} {1249} (\bibinfo {year} {2004})}\BibitemShut {NoStop}%
\bibitem [{\citenamefont {Mejia}(2015)}]{mejia2015very}%
  \BibitemOpen
  \bibfield  {author} {\bibinfo {author} {\bibfnamefont {J.~N.~Q.}\ \bibnamefont {Mejia}},\ }\href@noop {} {\emph {\bibinfo {title} {Very Nonlinear Quantum Optics}}}\ (\bibinfo  {publisher} {University of Toronto (Canada)},\ \bibinfo {year} {2015})\BibitemShut {NoStop}%
\bibitem [{\citenamefont {Panda}\ \emph {et~al.}(2019)\citenamefont {Panda}, \citenamefont {Palaniappan}, \citenamefont {Xiong}, \citenamefont {Bridgeford}, \citenamefont {Mehta}, \citenamefont {Shen},\ and\ \citenamefont {Vogelstein}}]{panda2019hyppo}%
  \BibitemOpen
  \bibfield  {author} {\bibinfo {author} {\bibfnamefont {S.}~\bibnamefont {Panda}}, \bibinfo {author} {\bibfnamefont {S.}~\bibnamefont {Palaniappan}}, \bibinfo {author} {\bibfnamefont {J.}~\bibnamefont {Xiong}}, \bibinfo {author} {\bibfnamefont {E.~W.}\ \bibnamefont {Bridgeford}}, \bibinfo {author} {\bibfnamefont {R.}~\bibnamefont {Mehta}}, \bibinfo {author} {\bibfnamefont {C.}~\bibnamefont {Shen}},\ and\ \bibinfo {author} {\bibfnamefont {J.~T.}\ \bibnamefont {Vogelstein}},\ }\bibfield  {title} {\bibinfo {title} {hyppo: A multivariate hypothesis testing python package},\ }\href@noop {} {\bibfield  {journal} {\bibinfo  {journal} {arXiv preprint arXiv:1907.02088}\ } (\bibinfo {year} {2019})}\BibitemShut {NoStop}%
\end{thebibliography}%

\end{document}